\documentclass[aps,groupedaddress,nofootinbib,11pt]{revtex4}

\usepackage[dvips]{epsfig}
\usepackage{amsfonts}
\usepackage{amsmath}
\usepackage{}
\usepackage{color}

\definecolor{Blue}{rgb}{0.3,0.3,0.9}
\definecolor{Red}{rgb}{1.0,0.0,0.0}
\definecolor{Green}{rgb}{0,0.4,0}
\definecolor{Violet}{rgb}{0.4,0.0,0.6}
\definecolor{Cyan}{rgb}{0.0,0.4,0.6}
\definecolor{Orange}{rgb}{1.0,0.4,0.0}

\newcommand{\Eqref}[1]{(\ref{#1})}
\newcommand{\ket}[1] {\mbox{$ \vert #1 \rangle $}}

\newcommand{\bra}[1] {\mbox{$ \langle #1 \vert $}}
\newcommand{\abs}[1] {\mbox{$ \vert #1 \vert $}}

\newcounter{subequation}[equation] \makeatletter
\expandafter\let\expandafter\reset@font\csname
reset@font\endcsname
\newenvironment{subeqnarray}
  {\arraycolsep1pt
    \def\@eqnnum\stepcounter##1{\stepcounter{subequation}{\reset@font\rm
      (\theequation\alph{subequation})}}\eqnarray}
  {\endeqnarray\stepcounter{equation}}
\makeatother

\newcommand{\ba}{\begin{eqnarray}}
\newcommand{\ea}{\end{eqnarray}}
\newcommand{\sba}{\begin{subeqnarray}}
\newcommand{\sea}{\end{subeqnarray}}

\begin{document}

\vskip 1truecm

 \title{Quantum corrections during inflation and conservation of adiabatic perturbations}
 \author{David Campo}
 \email[]{dcampo@astro.physik.uni-goettingen.de}
 \affiliation{Lehrstuhl f\"{u}r Astronomie,
Universit\"{a}t W\"{u}rzburg, Am Hubland,
D-97074 W\"{u}rzburg, Germany\\
and Georg-August-Universit\"{a}t, 
Institut f\"{u}r Astrophysik,
Friedrich-Hund-Platz 1, 37077 G\"{o}ttingen, Germany}
 \begin{abstract} 
The possibility that quantum corrections break the conservation of superhorizon 
adiabatic perturbations in single field inflation is examined.
I consider the lowest order corrections from massless matter fields 
in the Hamiltonian formalism.
Particular emphasis is therefore laid on the renormalization.
The counterterms are the same as in the Lagrangian formalism.
The renormalized value of the tadpole is zero.
I find a possible secular dependence of the power spectrum at 
one loop due to the trace anomaly,
but this result depends on the approximation of the modes
and is inconclusive.
The symmetry (not) violated by the quantum corrections is the invariance
by dilatation. 
Perspectives on the backreaction problem are briefly discussed. 
 \end{abstract}
\maketitle

Inflation would have probably remained at the stage of 
a beautiful theoretical idea if it had not been possible to 
relate cosmological perturbations between the inflationary period 
to the matter and radiation spectra observed today. 
This possibility is truly amazing given that it implies 
relating cosmological perturbations between two periods separated by 
a vast energy range about which we know very little 
(reheating, phase transitions, and of course inflation itself).
What saves us is the fact that the scales which are observed today 
where outside the horizon during most of the time that these phenomena took place,
and were decoupled from what was happening 
on much shorter scales
\footnote{By this I really mean that the mechanism at work is decoupling, and not acausality.
By definition of inflation, the scale of causality is the size of the 
horizon at the beginning of inflation, and is therefore 
(much) larger than any observable scale today if inflation is to be of any use.}.

This decoupling takes on the form of a conservation law.
It has long been known that 
certain gauge-invariant combinations of the first order metric and 
matter perturbations become constant outside the horizon \cite{BST}.
This is the case for instance of 
the variable 
$\zeta = - \psi + \frac{H}{\rho_0 + p_0} \delta u$ 
where $\psi$ parametrizes (the diagonal part of the) perturbations of the spatial metric, $H$ 
is the Hubble rate, $\rho_0$ and $p_0$ are the background energy density 
and pressure respectively, and $\delta u$ is the potential of energy flow $\delta T^0_{\,\, i} = \partial_i \delta u$. 
It is related to the intrinsic curvature of spatial surfaces by 
${\cal R} = -\frac{4}{a^2} \partial^2 \zeta$.
A mode of wave number $q$ follows the equation
\ba \label{ratezeta}
  \frac{d\zeta_q}{dN} = - 3 \left( \frac{d \rho_0}{dN}  \right)^{-1} \, \delta p_{\rm nad}
  + O\left( \frac{q^2}{a^2 H} \right)
\ea
where $N$ is the number of efolds, i.e. $a=\exp(N)$.
The first term 
(the non-adiabatic pressure $\delta p_{\rm nad} = \delta p - \frac{\dot p_0}{\dot \rho_0} \delta \rho$) 
vanishes for superhorizon modes in many relevant cases, including
single field inflation \cite{WeinbergNG}. The second term regroups gradients 
and vanishes exponentially fast once the mode exits the horizon.

Two derivations of \Eqref{ratezeta} are particularly interesting 
for their generality and the physical insight they bring. 
In the derivation offered in \cite{conservWands,SasakiNL}, 
equation \Eqref{ratezeta} and its nonlinear generalization descend directly from the 
conservation of energy projected on 
the hypersurface of constant time $n^b \nabla_a T^{a}_{\,\, \, b}=0$
(the vector $n^a$ is the unit normal to these hypersurfaces). 
In an alternative derivation \cite{Maldacena}, Maldacena showed that 
in the infinite wavelength limit, the value of the action on a solution of the
non-linear Hamilton constraint in single field inflation is a boundary term
\ba \label{proofMalda}
  S = -2 \int\!\!d^3xdt \, \partial_t\left( a^3 e^{3\zeta} H  \right) 
  +  O\left( \frac{q^2}{a^2} \right)
\ea 
It does not contribute to the non-linear classical equations of motion, which therefore
admit the solution $\zeta = {\rm cte} + O(q^2/a^2)$.
The physical contend of (\ref{proofMalda}) and its relation to the first derivation
of (\ref{ratezeta}) is better revealed 
if we use the number of efolds $N = \ln a$ as the time coordinate. 
One then notices that for $q=0$, the action \Eqref{proofMalda} 
changes by a total derivative when the coordinates are rescalled
\ba \label{symm q=0}
  N \mapsto N + \lambda \, , \qquad x^i \mapsto e^{-\lambda} x^i
\ea
Equation \Eqref{proofMalda} thus implies the invariance by dilatation
of the field equations at vanishing wavenumber, 
which is obviously a symmetry of (\ref{ratezeta}). 
\footnote{The fact that a rescaling of the conformal time corresponds 
to a translation of the variable $N$ makes it intuitively clear that
the two derivations are related, since invariance by translation
is associated with conservation of energy.
To be more precise, there exists an equivalence class of 
coordinate systems in which the metric
reads $ds^2 = a^2 e^{2\zeta} \left( - d\tilde \tau^2 + d{\bf x}^2 \right)$ 
up to corrections $O(q^2/a^2)$ \cite{SasakiNL,WeinbergNG}. 
This implies that the 
normal to the constant time hypersurfaces $n$ becomes aligned with 
the conformal Killing vector $\partial/\partial {\tilde \tau}$.}

So much for the classical theory. What can we expect in the quantum theory ?
A breakdown of the conservation law would be tight to a violation of the 
invariance by dilatation. Now,
under a Weyl transformation, i.e. a local scale transformation which leaves the coordinates 
fixed $g_{ab}(x) \mapsto e^{2 \Omega(x)} g_{ab}(x)$, the effecive action obtained 
by integrating the matter degrees of freedom changes according to
\ba \label{violation by trace}
  \delta_\Omega \Gamma = - \frac{1}{2} \int\!\!d^4x \, \sqrt{-g}\, \Omega \langle T^{a}_{\,\, \, a} \rangle
\ea
The trace of massless fields is fixed by the curvature,  
$\langle T^{a}_{\,\, a} \rangle = O(H^4)$. (This is true whether the
field is conformaly coupled or not, and by abuse of language 
I will call this term the trace anomaly regardless of the coupling to gravity.)
We can conclude that 
if non-conservation of $\zeta$ there is, it is 
because of the trace anomaly. 
This is only a heuristic argument. 
Whether this violation actually happens depends on
the details of the interactions (because the operator $T_{ab}$ is inserted into
loops and is therefore integrated over time with a certain weight function), and on the
details of the renormalization.
This is the calculation reported in this paper.

The rest of the paper is more technical.
The next section motivates this calculation and 
presents its technical aspects which are expanded in considerable details
in the subsequent five sections. 
The reader who does not want to get involved in the technicalities is invited to read 
the next section to get an idea of it and proceed to section \ref{sec:discussion}
where the main results are summarized and discussed.

\section{Introduction to the technical aspects of the calculation}
\label{sec:explain}

The conservation law is expressed as the time independence of the power spectrum
of superhorizon modes:
\ba \label{def power spectrum}
  {\cal P}(q,t) \equiv \int\!\!d^3x \, e^{-i{\bf q}{\bf x}} \langle \zeta(t,{\bf x})  \zeta(t,0) \rangle
  \to  \tilde {\cal P}(q)  \, , \qquad {\rm for} \quad \frac{q}{a H} \to 0
\ea

A first important step was achieved by Weinberg \cite{Weinberg,Weinberg2}.
Using the Hamiltonian formalism he showed  
that in a large class of theories, the late time contribution of 
quantum loops
\footnote{Unfortunately cosmologists use the word 'loop' to describe the
perturbative corrections in the so-called $\delta N$ formalism.
These corrections mix the quantum mechanical correction to 
a purely classical one which accounts for the passage from the 
flat gauge to the comoving gauge. These are not the loops considered here.} 
is at most powers of $\ln(a)$.
The proof of this theorem, which is also the way calculations in 
the Hamiltonian formalism have been done ever since,  
rests on the possibility to exchange the order of integrations, integrating first
over time at fixed external wavenumber $q$ and internal wavenumber $p$, and to take the limit
$t \to \infty$ before integrating over momentum.
The integral converges at early times because of the oscillatory
behaviour $\sim e^{-ip\tau}$ of the mode functions, and in the conditions of the theorem 
it also converges at late times because
the integrand contains sufficient factors of $1/a$.
The remaing integrals over the loop momenta are easely evaluated 
in dimensional regularization. 
This exchange is of course only allowed for convergent integrals,
which means that the ultraviolet divergences must be cancelled by counterterms. 
If such is the case, the integral over momentum
is effectively dominated by values $p\sim q$, and 
the value of the whole loop
is effectively set at $q\tau \simeq -1$
like tree graphs. 

An important point is that the $\ln(a)$ do not necessarily occur.
The proof of the theorem is algebraic (one counts the number of factors of $a$),
and the theorem only states that they can occur (as the limiting case).
Whether they do or not depends on the details of the interactions
and on the renormalization. 
To clarify these subtleties is one of the objectives of this work.

Why should one care about this, and why is it a non trivial problem ?
The first question is a legitimate one indeed. 
The possible violation of the conservation of $\zeta$ is certainly not an urgent problem
because if it occurs, it is likely to be too small to have observable consequences. 
My answer to the first question is that the calculation reported here 
should be considered as a baby step towards a better understanding
of the backreaction problem, in two respects: a theoretical one, 
with regard to the identification of proper observables 
to quantify the effect, and a pratical one, to develop and test 
the techniques to be used in future calculations of the backreaction.
(I will return to this in section \ref{sec:backreaction}.) 
Indeed, the Hamiltonian formalism lends itself particularly to calculations of 
the correlation functions of cosmological perturbations because
one can work with the physically relevant (non linear) variables, e.g. $\zeta$ and the 
two helicities of the graviton.
Follows a considerable simplification of the algebra compared to the 
Lagrangian formalism where calculations are carried out
with a $4\times 4$ tensor $h_{ab}$ (the 'pseudo-graviton') and
involve ghost fields.
The downside of adopting the Hamiltonian formalism 
is that by abandoning manifest Lorentz invariance, 
we are deprived of a powerful instrument to identify counterterms.
This is the answer to the second question.
But since we have very good reasons to use the Hamiltonian formalism,
we must try and make it work.

The task consists of the following steps which also delineate the 
plan of the paper.
First, the Hamiltonian is calculated up to fourth order.
I will ignore gravitons and I will only calculate the corrections from matter 
at one loop order. These assumptions considerably simplify the 
calculations which are already relatively tedious.
The part of the interaction Hamiltonian relevant for this calculation 
consists of two pieces, a three vertex $H_3 \sim \zeta \sigma \sigma$ and a 
four vertex $H_4 \sim \zeta \zeta \sigma \sigma$. 
The calculation of $H_4$, which is the first new result
of the paper, is a bit lengthy because 
the constraints must be solved at second order and the relation between
field derivatives and momenta is non trivial by virtue of the derivative
couplings between matter and gravity. The canonical quantization is detailed 
in sec. \ref{sec:can quantization}.

Before calculating the loop corrections I present in section \ref{sec:counterterms} 
the expressions of the counterterms quadratic in the curvature, namely 
the Weyl tensor $C^2 = C^{abcd}C_{abcd}$, the Euler density ${\rm E}_4$,
the square of the Ricci scalar and the Dalembertian of the Ricci scalar.

Section \ref{sec:tadpole} is devoted to the renormalization of the tadpole
$\langle \zeta \rangle$, which is tantamount to the renormalization of 
the energy-momentum tensor. It is shown that the renormalized 
tadpole can be consistently set to zero.

The corrections to the power spectrum 
from $H_3$ and $H_4$ are considered in sections  \ref{sec:three vertex} and 
sec. \ref{sec:corr H_4}. They are shown to be
renormalized by the same geometric counterterms as the tadpole.
Several appendices complete the presentation of these calculations.
The following five sections are very factual. The important results are 
discussed in sec. \ref{sec:discussion}.

\section{Canonical quantization} 
\label{sec:can quantization}

\subsection{Model and conventions}

I use the model of \cite{Weinberg}.
Matter consists of one inflaton field $\varphi(t,{\bf x})$ 
characterized by a classical potential $V(\varphi)$ 
and a vector field $\vec \sigma(t,{\bf x}) = {}^T\left( \sigma_1,\, ...  \right)$ 
in the defining representation of $O({\cal N})$, with no self-interactions.
$\varphi(t,{\bf x})$ and $\vec \sigma(t,{\bf x})$ are 
indirectly coupled through gravity. 
We work in the gauge where the inflaton is homogeneous
on each space-like hypersurface, 
\ba \label{deffoliation}
  \varphi(t,{\bf x}) = \varphi(t) \, ,
\ea
and where the parametrization of the spatial sections 
exhibits the physical degrees of freedom of the problem
\ba \label{defgauge}
  h_{ij} = a^2(t) e^{2\zeta} \left(\exp \gamma\right)_{ij} \, \, , \qquad 
   \gamma_{ii} = 0 \,,  \qquad \partial_i \gamma_{ij}= 0 \, .
\ea
where $a(t)$ is the background scale factor. 
The remaining metric elements are
\ba \label{g_00 and g_0i}
  g_{00} = -N^2 + h_{ij} N^i N^j   
\, \, , \qquad 
  g_{0i} = h_{ij}N^j 
\ea
where $N$ and $N^i$ are the lapse and shift functions. The comoving time is noted 
$\tau=\int dt/a(t)$.

The Lagrangian density is
\ba \label{L}
  {\cal L}&=& \frac{1}{2} \sqrt{-g} 
  \left(  R + \dot \varphi^2 - 2V(\varphi) + g^{ab} \partial_a \vec \sigma \partial_b \vec \sigma \right)
\nonumber \\
  &=& \frac{a^3 e^{3 \zeta}}{2} \left[ N {\cal R} - 2 N V(\bar \varphi) + 
  \frac{\dot{\varphi}^2}{N} + 
\frac{1}{N} \left( E_{i j}  E^{ij} -  E^2 \right)  \right] 
\nonumber \\
  && \, + \frac{a^3 e^{3 \zeta}}{2N} 
  \left( \dot{\vec \sigma} - N^i \partial_i \vec \sigma \right)^2
  - \frac{N a e^{\zeta}}{2} \left[\exp \left( -\gamma\right)\right]^{ij} \, 
  \partial_i \vec \sigma  \partial_j \vec \sigma  \,\, 
  +  {\rm \,\, total \,\, derivative}
\ea
The second form is the one appropriate for the canonical formulation \cite{MTW}.
The new terms appearing in this equation are: 
\ba
  E_{ij} = \frac{1}{2} \left( \partial_t h_{ij} - \nabla_i N_j - \nabla_j N_i \right)
\ea
related to the extrinsic curvature of constant time hypersurfaces
by $K_{ij}= N^{-1} E_{ij}$, its trace $E= h^{ij}E_{ij}$, 
and the Ricci scalar curvature of spacelike surfaces 
${\cal R} =  h^{ij}{\cal R}_{ij}$. 
I will ignore gravitons.
Since the $\sigma_i$ are essentially ${\cal N}$ copies of the same field,
I will be sloppy in my notations and often write $\sigma$ for $\vec \sigma$.

Before proceeding with the solution of the constraints and the canonical quantization, 
I introduce a few notations and conventions.
I will from now on write the scale factor $a(t) = e^\rho$
and the Hubble rate $\dot \rho$.
The first slow-roll parameter is defined by
$ \epsilon = - \frac{\ddot \rho}{\dot \rho^2} = \frac{\dot \varphi^2}{2\dot \rho^2}$.
To simplify the Fourier transform of the upcoming expressions, 
the spatial gradient $\partial_i = \frac{\partial}{\partial x^i}$ 
and Laplacian
are defined with respect to the flat metric $\delta_{ij}$. 
Hence $\partial^i =  \delta^{ij} \partial_j = \partial_i$,
and $\partial^2 = \sum_{i=1}^{3} \frac{\partial^2}{\partial (x^i)^2} = \delta^{ij} \partial_i\partial_j$.
The inverse of the Laplacian is noted $\partial^{-2}$.
I will ofen use the notation $\widetilde{d^3p}$ for the measure $\frac{d^3p}{(2\pi)^3}$.
Conventions for the curvature are those of the book of Misner, Thorne and Wheeler \cite{MTW}: 
The components of the Riemann tensor are 
$R^{a}_{\,\, bcd} = \partial_c \Gamma^{a}_{bd} - \partial_d \Gamma^{a}_{bc} + 
\Gamma^{a}_{ec} \Gamma^{e}_{bd} - \Gamma^{a}_{ed} \Gamma^{e}_{bc}$,
and the Ricci tensor is $R_{bd} = R^{a}_{\,\, bad}$.

The algebra will be done in full generality, but for the 
calculation of the integrals in closed form some approximate solution
of the mode equations must be chosen.
The linear equation of $\zeta$ is (see the quadratic part of the action \Eqref{S_2})
\ba \label{EOMzeta}
  \ddot \zeta + \left(3\dot \rho + \frac{\dot \epsilon}{\epsilon}  \right) \dot \zeta - \frac{1}{a^2} \, \partial^2 \zeta = 0 
\ea
The linear equation of a massless scalar field with coupling 
$\frac{\xi}{2}\sqrt{-g} R \sigma^2$ is
\ba \label{EOMxi}
 \ddot \sigma + 3\dot \rho  \dot \sigma + 12\xi \dot \rho^2 \sigma - \frac{1}{a^2} \, \partial^2 \sigma = 0 
\ea 
Following \cite{Weinberg}, 
I will use the quasi-de Sitter approximation around the time of horizon crossing.
The positive frequency solutions in the asymptotic past
(i.e. corresponding to the free Bunch Davies vacuum) are then
\ba \label{solzeta}
  \zeta_q(t) &=& \zeta_q^0 \left( 1+ i q\tau \right) e^{-iq\tau}
\, , \qquad \abs{\zeta_q^0}^2 = \frac{4\pi G }{\epsilon_q} \, \frac{H_q^2}{2 q^3}
\ea
For a minimally coupled scalar field  $\xi = 0$ we have
\ba \label{modemin}
  \sigma_q(t) &=& \sigma_q^0 \left( 1+ i q\tau \right) e^{-iq\tau}
\, , \qquad \abs{\sigma_q^0}^2 =  \frac{H_q^2}{2 q^3}
\ea
and in the case of a conformal coupling $\xi = 1/6$, 
\ba \label{modeconf}
  \sigma_q = \frac{\sigma_q^0}{a} e^{-iq\tau} 
 \, , \qquad   \vert \sigma_q^0 \vert^2 = \frac{1}{2q}
\ea
The normalization constants $\zeta_q^0$ and $\sigma_q^0$ are fixed 
by the canonical commutation relations.

I will discuss in due course the validity of this approximation,
namely in sections \ref{sec:tadpole}, \ref{sec:time integrals}
 and appendix \ref{app:time int red spectrum}.

\subsection{Solutions of the energy and momentum constraints}

The Hamiltonian and momentum constraints obtained 
by variation of \Eqref{L} w.r.t. $N$ and $N^i$ are respectively
\ba
  N^2\left({\cal R} - 2V - h^{ij} \partial_i \vec \sigma \partial_j \vec \sigma  \right) &=& 
   E_{i j}  E^{ij} -  E^2 + \dot \varphi^2 + \left( \dot{\vec \sigma} - N^i \partial_i \vec \sigma \right)^2
\\
  \nabla_i\left[ N^{-1} \left( E^{i}_{\,\,j} - \delta^{i}_{\,\,j} E  \right) \right] 
  &=& \frac{1}{N} \partial_j \vec \sigma (\dot{\vec \sigma} - N^i \partial_i \vec \sigma)
\ea
They are solved iteratively using the Ans\"{a}tze
\ba
  N = 1 + \alpha + \alpha_2 + ... \, , \qquad
  N^i &=& \partial_i (B + B _2 ) + w^i+ w^i_{2} + ...  \, , \qquad \partial_i w_{(2)}^i = 0 
\ea
I will write first order quantities without an index to
avoid an exponential number of occurrences of the number one.
Some expressions are simpler if we use the conformal time $\tau$ 
and I will go freely from $t$ to $\tau$. 
For the latter, it proves convenient to absorb the factor of $a$
from $dt = a d\tau$ into new parameters 
\ba 
  \omega = a B \,, \qquad  \widetilde{w}^i = a w^i 
\ea
so that we have
$h_{ij}N^j dt dx^i = a^2 e^{-2\zeta}(\partial_i B + w^i) dt dx^i = 
a^2 e^{-2\zeta}(\partial_i \omega + \widetilde{w}^i) d\tau dx^i$.

Only the first two orders are necessary to 
calculate the action up to fourth order \cite{Maldacena} (see footnote
\ref{foot:action}).
The solutions at first order are 
\sba \label{constrains1}
  \alpha &=& \frac{\dot \zeta}{\dot \rho} = \frac{\zeta'}{\rho'} \\
  B &=& -\frac{\zeta}{a^2 \dot \rho} + \epsilon \partial^{-2} \dot \zeta \, , \qquad 
   \omega = - \frac{\zeta}{\rho'} + \epsilon \partial^{-2} \zeta'
\\
  w^i &=& 0
\sea
They do not depend on $\sigma$ because the field has a vanishing v.e.v.,
and they do not depend on the inflaton fluctuations because 
we work in the foliation \Eqref{deffoliation}.
At second order, the lapse and shift receive separate contributions from
matter and first order metric perturbations,
\ba \label{alphag+alpham}
   \alpha_2 = \alpha_g + \alpha_m \, , \qquad B_2 = B_g + B_m \, , \qquad w_2^i = w_g^i + w_m^i 
\ea
The purely matter contributions are
\ba \label{alpha_m}
  \partial^2 \alpha_m &=& \frac{1}{2\dot \rho} \partial_j \left(\dot \sigma \partial_j \sigma \right)
\\
\label{B_m}
   \partial^2B_m &=& - \frac{1}{4\dot \rho} \left( \dot \sigma^2 + \frac{(\partial \sigma)^2}{a^2} \right)  
  - \frac{V}{\dot \rho}  \alpha_m
\\
\label{w_m}
  \partial^2w_m^j &=& - 2  \dot \sigma \partial_j \sigma + 4 \dot \rho \partial_j \alpha_m 
\ea
and the contributions from first order metric perturbations are
\ba \label{alpha_g}
  2 \dot \rho \partial^2 \alpha_g &=& 
 -2 \partial^2 \left(\partial_i \zeta \partial_i B \right) 
  + \partial_j\left\{ 
   \left(\partial_j \alpha \partial^2 B - \partial_i \alpha \partial_i\partial_j B  \right)
  + 3 \partial_i \zeta \partial_i\partial_j B - \partial_j \zeta \partial^2 B
  \right\}
\\
\label{w_g}
  -\frac{1}{2} \partial^2 w_g^j &=& - 2\partial_j \left[ \dot \rho \partial_j \alpha_g 
  + \left(\partial_i \zeta \partial_i B \right) \right]
\nonumber \\
  && \, 
+ \left(\partial_j \alpha \partial^2 B - \partial_i \alpha \partial_i\partial_j B  \right)
  + 3 \partial_i \zeta \partial_i\partial_j B - \partial_j \zeta \partial^2 B
\ea
and 
\ba \label{B_g}
  4\rho' \partial^2 \omega_g &=& - 4a^2 V\alpha_g - 6\epsilon{\zeta'}^2
  - 2(\partial \zeta)^2 + 8\zeta \partial^2 \zeta
  - 12 \rho' \partial_i\zeta \partial_i \omega +  
  \left[  (\partial^2 \omega)^2 - \partial_i \partial_j \omega  \partial_i \partial_j \omega  \right] 
\ea

\subsection{The action up to fourth order}

We can now expand the action (\ref{L}) up to fourth order
\footnote{\label{foot:action}I brielfy recall why the solutions of the constraints at third and fourth order are not needed
for this calculation \cite{Maldacena}. Upon expanding the action \Eqref{L}, one observes that
$\alpha_4$ multiplies the zeroth order energy constraint, and 
$\alpha_3$ multiplies the first order energy constraint, which are both assumed 
to hold; terms in $\nabla N$ of third and fourth order cancel; 
$\nabla_{(i} N_{j)} \nabla^{(i} N^{j)}$ starts at second order, but after 
integration by parts the second order term turns out to vanish.
The final form (\ref{S_2}-\ref{S_4}) is obtained after several integration by parts,
use of the constraints equations, and having dropped spatial divergences.}. 
The quadartic part is
\ba \label{S_2}
  S_2 &=& \frac{1}{2}\int\!\!a^3 \, \left\{ 2\epsilon \left( \dot \zeta^2 - \frac{(\partial \zeta)^2}{a^2}  \right) 
  + \left( \dot \sigma^2 - \frac{(\partial \sigma)^2}{a^2} \right) \right\}
\ea
The cubic interactions are 
\ba \label{S_3}
   \begin{split}
    S_3 = \frac{1}{2}\int\!\!a^3 \, &\Biggl\{ -2\epsilon \zeta\frac{(\partial \zeta)^2}{a^2} 
  - 2\epsilon \frac{\dot \zeta^3}{\dot \rho}
  - 4 \partial^2B \, \partial_iB \, \partial_i\zeta - 
  \frac{\dot \zeta}{\dot \rho}\left[ \partial_i\partial_j B  \,  \partial_i\partial_j B - (\partial^2 B)^2 \right]
\nonumber \\
      & \, + \left(3\zeta - \frac{\dot \zeta}{\dot \rho} \right) \dot \sigma^2 - 2 \dot \sigma  \, \partial_i B \,  \partial_i \sigma
  - \left(\zeta + \frac{\dot \zeta}{\dot \rho} \right)\frac{(\partial \sigma)^2}{a^2} \Biggr\}
   \end{split}
\ea
I substituted $\dot \zeta/\dot \rho$ for $\alpha$ but left the other 
parameters of the lapse and shift to simplify the expressions.
The first line of $S_3$ regroups the purely gravitational cubic interactions $\zeta \zeta \zeta$ 
(since $B$ is independent of $\sigma$)
and the second line regroups gravity-matter interactions $\zeta\sigma\sigma$.
This result is not new: 
$\zeta^3$ interactions were derived in \cite{Maldacena} and the matter-$\zeta$
interactions in \cite{Weinberg}.
The fourth order interactions are
\ba \label{S_4}
    \begin{split}
      S_4 =  \frac{1}{2}\int\!\!a^3 \, &\Biggl\{ 
     -\epsilon\zeta^2\frac{(\partial \zeta)^2}{a^2}
    + 2\epsilon  \frac{\dot \zeta^4}{\dot \rho^2} + 2V(\alpha_2)^2 
 \nonumber \\
      & \,+ \left[\partial_iw_2^j \, \partial_iw_2^j - 4 \partial^2 B \, \partial_i B_2  \, \partial_i\zeta -
     4 \partial^2 B_2  \, \partial_i B  \, \partial_i\zeta - 6(\partial_i B  \, \partial_i\zeta)^2\right]
 \nonumber \\
     & \, + \left(3\zeta - \frac{\dot \zeta}{\dot \rho} \right) \left[ 
     2  \partial_i\partial_j B   \, \partial_i\partial_j B_2 - 2\partial^2 B  \, \partial^2 B_2 
     + 2 \partial_i\partial_j B \, \partial_iw_2^j - 4 \partial^2 B  \, \partial_i B  \, \partial_i\zeta \right]
 \nonumber \\
     & \, -\frac{\dot \zeta}{\dot \rho}\left(3\zeta - \frac{\dot \zeta}{\dot \rho} \right) 
     \left[ \partial_i\partial_j B  \,  \partial_i\partial_j B - (\partial^2 B)^2 \right]
 \nonumber \\
     & \,+ \left(\frac{\dot \zeta^2}{\dot \rho^2} + \frac{9}{2}\zeta^2 - 3\zeta\frac{\dot \zeta}{\dot \rho}  \right) \dot \sigma^2
      - 2\left(3\zeta - \frac{\dot \zeta}{\dot \rho} \right) \dot \sigma \partial_i B  \, \partial_i \sigma
 \nonumber \\
     & \, - 2\dot \sigma \partial_i B_2  \, \partial_i \sigma - 2\dot \sigma w_2^i  \, \partial_i \sigma 
     + (\partial_i B  \, \partial_i\sigma)^2 -
     \left(\frac{\zeta^2}{2} + \zeta\frac{\dot \zeta}{\dot \rho}  \right) \frac{(\partial \sigma)^2}{a^2}
 \Biggr\}
    \end{split}
\ea
The pure $\zeta$ part was previously given in \cite{Slothzeta^4}.
\footnote{For comparison with this reference, 
note the use of different conventions:
$\partial^2 = e^{-2\zeta}\delta_{ij} \partial_i \partial_j$ 
and $(\partial \zeta)^2 = e^{-2\zeta}\delta_{ij}   \partial_i \zeta \partial_j\zeta$.}

Only the Hamiltonian of the $\zeta \sigma \sigma$ interactions has 
been previously calculated \cite{Weinberg}.
I present the derivation of the complete Hamiltonian at fourth order in 
sec. \ref{sec:can quantization}. Before this, I check
the expression of the matter-$\zeta$ interactions
against the action given by the covariant formalism. In terms of 
the pseudo graviton tensor $h_{ab}$ defined by
\ba
  ds^2 = a^2(\tau) \left( \eta_{ab} + h_{ab}   \right) dx^a dx^b
\ea
the action up to fourth order is
\ba \label{covS}
  S = -\frac{1}{2} \int\!\!d^4x \, \sqrt{-g} \, 
  g^{ab} \partial_a \sigma \partial_b \sigma 
  =  -\frac{1}{2} \int\!\!d^4x \, a^4(\tau) \, 
  \left( \bar h^{ab} + H^{ab} \right) \partial_a \sigma \partial_b \sigma 
\ea
where the two auxilliary tensors are
\ba \label{bar h and H}
  \bar h_{ab} &=& h_{ab} - \frac{h}{2}\eta_{ab} \, , \qquad h = \eta^{cd} h_{cd}
\nonumber  \\
  H^{ab} &=& h_{ac}h^{c}_{\,\, b} - \frac{1}{2} h h_{ab} + 
  \frac{1}{4}\left( \frac{h^2}{4} - h_{cd}h^{cd} \right)\eta_{ab}
\ea
Their indices are raised and lowered with $\eta_{ab}$.
Substitution of \Eqref{defgauge} and the solutions of the constrains into 
\Eqref{bar h and H} gives
\ba \label{S_3cov}
  S_3^{\rm cov} &=& \frac{1}{2}\int\!\!a^3 \, 
  \left\{ \left( 3\zeta - \frac{\dot \zeta}{\dot \rho} \right) \dot \sigma^2 
  - 2\partial_i B \dot \sigma \partial_i \sigma - 
  \left(\zeta + \frac{\dot \zeta}{\dot \rho} \right)  \frac{(\partial \sigma)^2}{a^2} 
  \right\}
\ea
which is equal to the second line of \Eqref{S_3}, and
\ba \label{S_4cov}
  \begin{split}
     S_4^{\rm cov} = \frac{1}{2}\int\!\!a^3 \, 
   &\Biggl\{
    - \alpha_2\left( \dot \sigma^2 + \frac{(\partial \sigma)^2}{a^2} \right)
  + \left(\frac{\dot \zeta^2}{\dot \rho^2} + \frac{9}{2}\zeta^2 - 3\zeta\frac{\dot \zeta}{\dot \rho}  \right) \dot \sigma^2
  - 2\left(3\zeta - \frac{\dot \zeta}{\dot \rho} \right) \dot \sigma \partial_i B \partial_i \sigma
\nonumber \\
  & \, - 2\dot \sigma \partial_i B_2 \partial_i \sigma - 2\dot \sigma w^i \partial_i \sigma 
  + (\partial_i B \partial_i\sigma)^2 -
  \left(\frac{\zeta^2}{2} + \zeta\frac{\dot \zeta}{\dot \rho}  \right) \frac{(\partial \sigma)^2}{a^2}  
  \Biggr\}
  \end{split}
\ea
But for the first term proportional to $\alpha_2$, this is the last two lines of
\Eqref{S_4}. There is actually no discrepancy between the two results:
the terms $- \alpha_2\left( \dot \sigma^2 + \frac{(\partial \sigma)^2}{a^2} \right)$ 
in \Eqref{S_4cov} can be combined with the other terms proportional to $\alpha_2$
and simplified using the second order constraint equation to yield 
the term $2V (\alpha_2)^2$ that appears in \Eqref{S_4}.

\subsection{Canonical quantization}
\label{sec:can quantization}

I now proceed with the canonical quantization.
One first calculates the conjugate momenta 
\ba
  \pi_\zeta = \frac{\delta S}{\delta \dot \zeta} \, , \qquad  
  \pi_\sigma = \frac{\delta S}{\delta \dot \sigma}
\ea
and inverts these expressions. 
Schematically we have
\ba \label{dotzeta(pi)}
  \dot \zeta &=& \dot \zeta(\pi_\zeta,\zeta, \pi_\sigma, \sigma) = \pi_\zeta + 
 \delta_2 \dot \zeta + \delta_3 \dot \zeta + ...
\\
  \dot \sigma &=& \dot \sigma(\pi_\zeta,\zeta, \pi_\sigma, \sigma)= \pi_\sigma + 
 \delta_2 \dot \sigma + \delta_3 \dot \sigma + ...
\ea 
where $\delta_{2}$ (resp. $\delta_{3}$) regroups the terms second (resp. third) order 
in the field variables, and the dots represent higher order terms.
To simplify the expressions, 
I do not write the factors of $a^3$ and $\epsilon$ associated with each factor of $\pi$. 
One then substitutes these expressions into
\ba \label{defH}
  H = \int\!\!d^3x \, \left\{  \pi_\zeta \dot \zeta + \pi_\sigma\dot \sigma - 
  \left( {\cal L}_2 + {\cal L}_3 + {\cal L}_4 + ... \right) \right\}
\ea
In the fourth order Lagrangian ${\cal L}_4$ we can directly replace 
$\dot \zeta$ by $\pi_\zeta$ and $\dot \sigma$ by $\pi_\sigma$.
The substitution in ${\cal L}_3$ requires the expressions \Eqref{dotzeta(pi)}
up to second order in the field variables, and 
we {\it a priori} need these relations up to third order
to substitute them into $\pi_\zeta \dot \zeta + \pi_\sigma\dot \sigma - {\cal L}_2$.
It turns out that these contributions cancel:
\ba \label{cancelation}
  \pi_\zeta \dot \zeta  - {\cal L}_2(\zeta,\dot \zeta) &=&
  \pi_\zeta \left( \pi_\zeta + \delta_2 \dot \zeta + \delta_3 \dot \zeta \right) 
  - \frac{1}{2}\left[ \pi_\zeta^2 + 2 \pi_\zeta \delta_2 \dot \zeta + 2 \pi_\zeta\delta_3 \dot \zeta 
  + ( \delta_2 \dot \zeta)^2 + ... \right]
 \nonumber \\
  &=& \frac{1}{2}\left[ \pi_\zeta^2 - ( \delta_2 \dot \zeta)^2 + ...\right]
\ea
and similarly for the $\sigma$-sector. The dots stand for higher order terms.
Thus we only need to calculate the momenta at second order, a most welcome simplification.

Because of the non-local form of the interactions, 
it is easier to proceed with the Fourier representation of the action.
With the convention that repeated indices are summed, the cubic action has the form 
\ba \label{defc,d,...}
  S_3 &=& \frac{1}{2} \delta({\bf p}+{\bf k}+{\bf l}) \left\{ 
   c_{{\bf p},{\bf k},{\bf l}} \dot \zeta_{\bf p} \dot \zeta_{\bf k}\dot \zeta_{\bf l} 
  + d_{{\bf p},{\bf k},{\bf l}} \dot \zeta_{\bf p} \dot \zeta_{\bf k} \zeta_{\bf l} 
  + e_{{\bf p},{\bf k},{\bf l}} \dot \zeta_{\bf p} \zeta_{\bf k} \zeta_{\bf l} \right\}
\nonumber \\
  &+&\frac{1}{2} \delta({\bf p}+{\bf k}+{\bf l}) \left\{ 
  \left(3\zeta_{\bf p} - \frac{\dot \zeta_{\bf p}}{\dot \rho} \right) \dot \sigma_{\bf k}\dot \sigma_{\bf l}  
 + f_{{\bf p},{\bf k},{\bf l}} \dot \sigma_{\bf p}\sigma_{\bf k} \zeta_{\bf l} 
 + g_{{\bf p},{\bf k},{\bf l}} \dot \sigma_{\bf p}\sigma_{\bf k} \dot \zeta_{\bf l}
 + h_{{\bf p},{\bf k},{\bf l}} \dot \zeta_{\bf p} \sigma_{\bf k} \sigma_{\bf l}   
  \right\} + ... \qquad
\ea
The dots stand for the terms without time derivatives.
The expressions of $c,d,...$ are read off \Eqref{S_3}:
\ba \label{def c,d,...}
  c_{{\bf p},{\bf k},{\bf l}} &=& - \frac{\epsilon}{\dot \rho} \left( 2 + 
  \epsilon \frac{\sigma({\bf k},{\bf l})}{{\bf k}^2 \, {\bf l}^2} \right)
\nonumber \\
  d_{{\bf p},{\bf k},{\bf l}} &=& -2 \epsilon\left( \frac{1}{a^2 \dot \rho} \frac{\sigma({\bf k},{\bf l})}{{\bf k}^2 } 
  + 2 \epsilon \frac{{\bf k}\cdot {\bf l}}{{\bf k}^2} \right) = 2\epsilon \, \tilde d_{{\bf p},{\bf k},{\bf l}}
\nonumber \\
  e_{{\bf p},{\bf k},{\bf l}} &=& - \frac{1}{a^2 \dot \rho} \left(  \frac{\sigma({\bf k},{\bf l})}{(a\dot \rho)^2 } 
  + 4 \epsilon \frac{{\bf k}^2}{{\bf p}^2} {\bf p} \cdot {\bf k}    \right)
\nonumber \\
  f_{{\bf p},{\bf k},{\bf l}} &=& -2  \frac{ {\bf k}\cdot {\bf l} }{a^2 \dot \rho} \, , \qquad 
  g_{{\bf p},{\bf k},{\bf l}} = -2\epsilon \frac{{\bf k}\cdot{\bf l}}{{\bf l}^2 } \, , \qquad 
  h_{{\bf p},{\bf k},{\bf l}} =  \frac{ {\bf k}\cdot {\bf l} }{a^2 \dot \rho}
\ea
and the auxiliary function is 
\ba
  \sigma({\bf k},{\bf l}) = \left( {\bf k}\cdot {\bf l} \right)^2- {\bf k}^2 \, {\bf l}^2
\ea
Note that these vertices have not been symmetrized, so that it is important
to respect the ordering of the momenta according to the definitions \Eqref{defc,d,...}
of the functions $c,d, ...$.

We will later need the value of these coefficients for ${\bf p} = 0, {\bf k} = - {\bf l}$.
Most of the terms above are singular because they originate from the inversion of 
the solution $\partial^2 B = ...$, and this equation is not defined for ${\bf p} = 0$.
This quandary is resolved once one remembers that the zero mode of $\zeta$
is a gauge mode and as such does not posses a well defined momentum.
So we must for consistency take the value of these coefficients to be zero.

Taking the variations of (\ref{defc,d,...}) 
with respect to $\dot \zeta_{\bf q}$ and $\dot \sigma_{\bf q}$ and inverting the
expressions up to second order in the field variables, one gets
\ba \label{delta2pisigma}
  \delta_2 \dot \sigma_{-{\bf q}} &=& - \frac{1}{2}\delta({\bf q}+{\bf p}+{\bf k}) \left\{ 
  2\left(3\zeta_{\bf p} - \frac{\dot \zeta_{\bf p}}{\dot \rho} \right) \pi^\sigma_{-{\bf k}}  
  +  f_{{\bf q},{\bf p},{\bf k}}\sigma_{\bf p} \zeta_{\bf k} 
  +  g_{{\bf q},{\bf p},{\bf k}}\sigma_{\bf p} \pi^\zeta_{-{\bf k}}    
  \right\}
\\ 
 \label{delta2pizeta}
  2\epsilon \, \delta_2 \dot \zeta_{-{\bf q}} &=& - \frac{1}{2}\delta({\bf q}+{\bf p}+{\bf k}) \left\{ 
  C_{{\bf q},{\bf p},{\bf k}}
  \pi^\zeta_{-{\bf p}} \pi^\zeta_{-{\bf k}}
  \tilde D_{{\bf q},{\bf p},{\bf k}} \pi^\zeta_{-{\bf p}}  \zeta_{\bf k} 
 + e_{{\bf q},{\bf p},{\bf k}} \zeta_{\bf p} \zeta_{\bf k} 
  \right\}
\nonumber \\
  &&- \frac{1}{2}\delta({\bf q}+{\bf p}+{\bf k}) \left\{  
  - \frac{1}{\dot \rho} \pi^\sigma_{-{\bf p}} \pi^\sigma_{-{\bf k}} 
  + g_{{\bf p},{\bf k},{\bf q}} \pi^\sigma_{-{\bf p}} \sigma_{\bf k} 
  + h_{{\bf q},{\bf p},{\bf k}} \sigma_{\bf p} \sigma_{\bf k}   
  \right\}
\ea
where 
\ba
  C_{{\bf q},{\bf p},{\bf k}} &=& c_{{\bf q},{\bf p},{\bf k}} + c_{{\bf p},{\bf q},{\bf k}} + c_{{\bf p},{\bf k},{\bf q}}
\nonumber \\
  \tilde D_{{\bf q},{\bf p},{\bf k}} &=& \tilde d_{{\bf q},{\bf p},{\bf k}}+ \tilde d_{{\bf p},{\bf q},{\bf k}} 
\ea
I recall that to simplify the expressions, I have not written the factors of $a^3$ and $\epsilon$ on the r.h.s.
The expression of $\delta_2 \dot \sigma$ has a simple form in the position representation
\ba
   \delta_2 \dot \sigma(x) = - \left( 3 \zeta - \frac{\dot \zeta}{\dot \rho} \right) \pi_\sigma + N^i \partial_i \sigma
\ea
which could have been obtained directly from \Eqref{L}. 
The expression of $\delta_2 \dot \zeta(x)$ is on the other hand far less neat-looking
and I do not write it.
Substitution of these expressions into \Eqref{defH} and using the expressions 
(\ref{delta2pisigma}-\ref{delta2pizeta}) to simplify some terms yields
\ba 
\label{H_2}
  H_2 &=& \frac{a^3}{2}\int\!\!d^3x \, 2\epsilon \left\{  \left(\frac{\pi_\zeta}{2\epsilon a^3} \right)^2 
  + \frac{(\partial \zeta)^2}{a^2} \right\} + 
   \left\{  \left(\frac{\pi_\sigma}{a^3} \right)^2 + \frac{(\partial \sigma)^2}{a^2} \right\}
\\
\label{H_3compact}
  H_3 &=& - \int\!\!d^3x \,{\cal L}_3
\\
\label{H_4compact}
  H_4 &=& \int\!\!d^3x \, \left\{ - {\cal L}_4 
  + \frac{a^3}{2}\left[ 2\epsilon (\delta_2 \dot \zeta)^2 +   ( \delta_2 \dot \sigma)^2 \right] \right\}
\ea
The $+$ sign in front of the last two terms of \Eqref{H_4} is not a typo. It is the sum of 
$ - \frac{1}{2} (\delta_2 \dot \zeta)^2$ from \Eqref{cancelation} and $+(\delta_2 \dot \zeta)^2$
obtained by collecting the terms produced by the variation of ${\cal L}_3$ in the substitution
$\dot \zeta \mapsto \pi + \delta_2 \dot \zeta$ (by the very definition of $\delta_2 \dot \zeta$).
These last two terms are produced by the inversion of the relation $\pi(\dot \zeta)$ and are 
therefore peculiar to the canonical quantization of theories with derivative couplings.

The procedure of canonical quantization is completed after replacing $(\zeta, \, \pi)$ in the above expressions 
by the interacting picture variables $(\zeta_I, \, \pi_I)$. Their evolution is by definition governed 
by $H_2$, and the interaction Hamiltonian is given by (\ref{H_3compact}) and (\ref{H_4compact}) with the
replacement
\ba
   \pi_\zeta \mapsto a^3 \epsilon \, \dot \zeta_I  \, \, , \qquad 
  \pi_\sigma \mapsto a^3  \dot \sigma_I 
\ea
In the following I do not write the indices $I$.
I only write the part of the interacting Hamiltonian 
relevant for the calculation of matter loops:
\ba \label{H_3}
  H_3 &=&-\frac{a^3}{2} \int\!\!d^3x \, 
   \left\{\left(3\zeta - \frac{\dot \zeta}{\dot \rho} \right) \dot \sigma^2 - 2 \dot \sigma \partial_i B \partial_i \sigma
  - \left(\zeta + \frac{\dot \zeta}{\dot \rho} \right)\frac{(\partial \sigma)^2}{a^2} \right\}
\ea
and
\ba \label{H_4}
  H_{4} &=&  \frac{a^3}{2} \int\!\!d^3x \, 
  \left\{\left(\frac{9}{2}\zeta^2 - 3\zeta\frac{\dot \zeta}{\dot \rho}  \right) \dot \sigma^2
  +\left(\frac{\zeta^2}{2} + \zeta\frac{\dot \zeta}{\dot \rho}  \right) \frac{(\partial \sigma)^2}{a^2}  \right\} 
  + H^{\rm \alpha,B}_{4} + H^{\rm extra}_{4}  
\ea
The first vertex is the sum of the relevant part of $-{\cal L}_4$ and $\frac{a^3}{2} ( \delta_2 \dot \sigma)^2 $.
The second term stems from having fixed the gauge and 
solved the constrains 
\ba \label{H_4constrains}
   H^{\rm \alpha,B}_{4} &=& -\frac{a^3}{2} \int\!\!d^3x \, \, 4V\alpha_g \alpha_m 
  + \left[2 \partial_iw_m^j \, \partial_iw_g^j - 4 \partial^2 B \, \partial_i B_m  \, \partial_i\zeta -
 4 \partial^2 B_m \, \partial_i B \, \partial_i \zeta \right]
\nonumber \\
  &&\qquad \quad + \left(3\zeta - \frac{\dot \zeta}{\dot \rho} \right) \left[ 
 2  \partial_i\partial_j B \, \partial_i\partial_j B_m - 2\partial^2 B \, \partial^2 B_m 
 + 2 \partial_i\partial_j B \, \partial_iw_m^j \right]
\ea
The third term in \Eqref{H_4} comes from the inversion of the relation between
time derivatives of the $\dot \zeta$ and $\pi_\zeta$, 
i.e. the last two terms in \Eqref{H_4compact}. 
In the momentum representation, it is given by 
\ba \label{H_4extra}
  H^{\rm extra}_{4} &=& - \frac{a^3}{8\epsilon \dot \rho} \int \delta({\bf q} + {\bf l} + {\bf l}') \, 
  \left[ \dot \sigma_{\bf l}  \dot \sigma_{{\bf l}'} - 
  \frac{{\bf l} \cdot {\bf l}'}{a^2} \sigma_{\bf l}  \sigma_{{\bf l}'} 
  - 2\epsilon \frac{{\bf l}' \cdot {\bf q}}{{\bf q}^2} \dot \sigma_{\bf l}  \sigma_{{\bf l}'} \right]
\nonumber \\ 
 && \qquad \times \, \delta( {\bf p} + {\bf k} - {\bf l} - {\bf l}') 
 \left[ C_{{\bf q},{\bf p},{\bf k}} \dot \zeta_{\bf p}  \dot \zeta_{\bf k} 
  + \tilde D_{{\bf q},{\bf p},{\bf k}} \dot \zeta_{\bf p}  \zeta_{\bf k} + 
 \frac{1}{2}\left( e_{{\bf q},{\bf p},{\bf k}} + e_{-{\bf q},{\bf p},{\bf k}} \right) \zeta_{\bf p}  \zeta_{\bf k} \right]
\ea
These additional interactions are the price to pay for the 
tremendous reduction of the number of variables
compared to the covariant formalism. One way or the other,
appearance of such terms is inevitable because we need to fix a gauge in order to 
quantize the theory, which in the covariant formalism comes at the price of ghost fields. 

The basic observation one can make at this point is that
the passage from \Eqref{S_4cov} to \Eqref{H_4} generated a certain number
of new terms. It is thus not obvious that the ensuing divergences 
can be absorbed by local geometrical counterterms  
as in the Lagrangian formalism.
It turns out however that both $H^{\rm \alpha,B}_{4}$ and $H^{\rm extra}_{4}$
have a vanishing contribution at one loop.

\section{Geometric counterterms}
\label{sec:counterterms}

When I carried out the canonical quantization
I did not include the counterterms in the action.
I justify this choice by refering to the point of view of effective theories, 
which implies a consistent perturbative treatment \cite{woodard highderiv,Simon}. 
Because the effective action is a perturbative expansion in $\hbar$,
counterterms of dimension $\left[ {\rm mass} \right]^{n}$ with $n > 4$
produced order by order by the loop expansion must 
not be considered as part of the classical action, but only as additional
vertices.
This means in particular that they must be discarded from the propagator
(in the Lagrangian formalism) and from the canonical momenta (in the Hamiltonian formalism),
despite that they contain terms linear and quadratic in $\zeta$.
I will therefore take $\int dt \,H_{\rm cnt} = - S_{\rm cnt}$.

The expressions of the Weyl tensor to first order and of the 
Ricci tensor and Ricci scalar to second order 
can be found in appendix \ref{app:curvature}.

\subsection{Squared Weyl tensor}

Let us begin with the squared Weyl tensor,
\ba \label{S_W}
  S_{\rm W} &=& \frac{C_{\rm W}}{2} \int\!\!d^4x \sqrt{-g} \,  C^2
=   \frac{C_{\rm W}}{2}\int\!\!d\tau d^3x \, 32(\epsilon \rho' \zeta')^2 + O(\zeta^3)  
\ea
As I explained I posit that the counterterm in the Hamiltonian formalism is 
\ba  \label{H_W}
  H_{\rm W}  = -16 C_{\rm W}  (\epsilon \rho')^2  
 \int\!\!\frac{d^3 p}{(2\pi)^3} \, \zeta'_{\bf p} \zeta'_{-{\bf p}}
\ea
With the notation $Q(\tau) = \zeta_{\bf q}(\tau) \zeta_{-{\bf q}}(\tau)$, the corresponding
correction to the power spectrum is
\ba  \label{Delta Weyl}
  \Delta_{\rm W} {\cal P} &=& - 2 Im \int^{\tau}_{-\infty}\!\!d\tau_1 \, \langle H_{\rm W}(\tau_1) Q(\tau) \rangle 
\nonumber \\
  &=&  64 C_{\rm W} \epsilon^2  Im\left\{ \int_{-\infty}^{\tau}\!\!d\tau_1 \, 
   \left[\rho'  \zeta_q'(\tau_1)\zeta_q^*(\tau) \right]^2 \right\}
\nonumber \\
   &=& 32 C_{\rm W} \left( \epsilon \vert \zeta_q^0 \vert^2 \right)^2 q^3 (1-q^2\tau^2)
\ea
In sec. \ref{sec:three vertex} 
this term will be shown to absorbe the divergences of the 
corrections from the interaction $\zeta \sigma\sigma$.

\subsection{Squared Ricci scalar}
\label{sec:cnt R^2}

The other counterterms are more involved because the background is not
Ricci flat, so they must be calculated at second order. The simplest 
is to start with the Ricci scalar squared. Let me note 
$\tilde R = 6\frac{a''}{a} = 6{\rho'}^2(2-\epsilon)$. We have,
\ba \label{counterterm R^2}
    S_{R^2} &=& \frac{C_{\rm R^2}}{2} \int\!\!d^4x \sqrt{-g} R^2 
 \\
   &=& \frac{C_{\rm R^2}}{2}\int\!\! \tilde R^2 \left( 1 + 3\zeta + \alpha + \frac{9}{2} \zeta^2 + 3\zeta \alpha  \right)
  + 8 \tilde R \epsilon \rho' \zeta'
  + \tilde R\left(  (2+\epsilon) T_{00} + \frac{T_{00}'}{\rho'}\right)
\nonumber \\
  && \qquad \,\,\, + \, 4\tilde R \left(\frac{1}{{\rho'}^2} (\partial^2 \zeta)^2 -  \alpha \partial^2 \alpha 
    + (1 + 5\epsilon + \epsilon_2) \, \alpha \partial^2 \zeta 
    - (3+2\epsilon) \, \zeta \partial^2 \zeta
  \right)
\nonumber \\
   && \qquad  \,\,\, + \, 432 {\rho'}^3 \zeta \zeta' 
   + 624 \epsilon (\rho' \zeta')^2 - 296 (\epsilon \rho' \zeta')^2 
\nonumber 
\ea
Here, $T_{00} = (\sigma')^2 + (\partial \sigma)^2$.
I neglected terms second order in the slow roll parameters $\epsilon = 1- \frac{\rho''}{(\rho')^2}$
and $\epsilon_2 = \frac{\epsilon'}{\rho'\epsilon}$.
Let me comment the various terms appearing in this expression: \\
1) The list of quadratic terms is exhaustive: no other term quadratic in $\zeta$ with 
four derivatives, including a maximum of two time derivatives of $\zeta$, can be constructed.
Therefore, the other counterterms
$R^{abcd}R_{abcd}$, $R^{a}_{\,\,b} R^{b}_{\,\,a}$ and $\Box R$ contain the same list of 
terms, only the prefactors differ.
This is checked explicitly in appendix \ref{app:curvature}. \\
2) One of the quadratic terms in \Eqref{counterterm R^2} is 
$(\epsilon \rho' \zeta')^2$, like $C^2$. There is therefore 
a degeneracy between those two counterterms that cannot be lifted from the sole term 
$\langle H_A H_A \zeta \zeta \rangle$ of sec. \ref{sec:three vertex}.\\
3) Appear also terms quadratic in the matter fields. They originate from the solution of the 
shift vector \Eqref{B_m}, $\partial^2 \omega_m = - \frac{T_{00}}{4\rho'} + ...$. 
They give a vanishing correction at lowest order
\ba
  2 Im\int^t\!\!d\tau_1\,  \langle H_{\rm cnt} \zeta_{\bf q}(\tau)  \zeta_{-{\bf q}}(\tau) \rangle = 
  12\vert \zeta_q(\tau) \vert^2 Im\int^\tau\!\!d\tau_1\, \frac{a''}{a}(\tau_1)
  \langle -4 (\sigma')^2 + (2+\epsilon) T_{00}  \rangle = 0
\ea
4) The quadratic terms $\tilde R^2 \left( \frac{9}{2} \zeta^2 + 3\zeta \alpha  \right)$
come from the expansion of $\sqrt{-g}$. They give a vanishing correction.\\
5) The terms ${\rho'}^3 \zeta \zeta'$ give a correction
\ba \label{cnt zeta zeta'}
  \Delta {\cal P} \propto 
   - 2  \, Im \int^\tau\!\!d\tau_1\, {\rho'}^3  \langle  \zeta_1 \zeta_1' \zeta^2(\tau) \rangle
  &=&  4  q^2 \, Im \left\{ \left(\zeta_q^0 \zeta_q^*(\tau) \right)^2 \, 
  \int^\tau\!\!\frac{d\tau_1}{\tau_1^2}\, (1+iq\tau_1) e^{-i2q\tau_1}   \right\}
\nonumber \\
  &=& - 4  q^3 \vert \zeta_q^0 \vert^4 \left\{ \ln(-q\tau) + {\rm cte} + O(q\tau) \right\}
\ea
I used the asymptotic behaviour of the exponential integral 
\ba \label{exp int}
  \int^{\tau}\!\!\frac{d\tau_1}{\tau_1} \, e^{-iq\tau_1} = \ln(-q\tau) + C + O(q\tau)
\ea
It is the only quadratic term of \Eqref{counterterm R^2} with this logarithmic dependence.
It will find its use in section \ref{sec:H_4}. 
The other quadratic terms of (\ref{cnt R^2}) containing ${\zeta'}^2$ or gradients 
give the same polynomial corrections:
\ba \label{polynome cnt}
 \int\!\! (\rho' \zeta')^2  \, , \quad 
 \int\!\! \frac{a''}{a}\zeta \partial^2 \zeta \, , \quad 
 \int\!\! \frac{a''}{a} \, \alpha \partial^2 \zeta  \, , \quad 
 \int\!\!  \frac{a''}{a{\rho'}^2} \, (\partial^2 \zeta)^2   \, , \quad 
 \int\!\! \frac{a''}{a}\alpha \partial^2 \alpha  \quad &\mapsto&  q^3 \vert \zeta_q^0 \vert^4 
\ea
These terms will be used to renormalize 
the corrections of the fourth order interactions.\\
6) Finally there are terms linear in $\zeta$. They appeared because the background is not Ricci flat. 
They serve to renormalize the tadpole $\langle \zeta \rangle$.
As we will need other counterterms in addition to the Ricci scalar squared, let me
calculate the contribution of a general linear counterterm
\ba \label{lin cnt}
 H_{\rm lin} = - \frac{1}{2 \tau^4} \int\!\!d^3x \left( 3 \gamma \, \zeta + \delta \, \frac{\zeta'}{\rho'} \right)
\ea
where $\gamma$ and $\delta$ are constants.
This expression also includes the linear terms from counterterms 
parametrized by a cosmological constant
$\Lambda \int\sqrt{-g}$ (or more generally of the inflaton potential) 
and a shift of the Newton constant $\int\sqrt{-g} R$.
For these terms as well as
for the squared Ricci scalar, the Euler density 
${\rm E}_4 = R^{abcd}R_{abcd} - 4 R^{ab}R_{ab} + R^2 $, and $\Box R$, their values are
\ba \label{list delta gamma}
  \Lambda &\rightarrow& \quad \gamma_{\Lambda} = \frac{1}{3\dot \rho^2} \, , 
  \qquad \qquad \,\,\, \,\,\, \,\,\delta_{\Lambda} = \frac{1}{3\dot \rho^2} 
\nonumber \\
  G_N  &\rightarrow& \quad \gamma_{G_N} = \frac{6}{\dot \rho^{2}}(2-\epsilon) \, , \qquad \, 
  \delta_{G_N} = \frac{2}{\dot \rho^{2}}(6-\epsilon)
\nonumber \\
  R^2 &\rightarrow& \quad \gamma_{R^2} = 36(2-\epsilon)^2 \, , \qquad 
  \, \delta_{R^2} = 12(12 + 4 \epsilon - \epsilon^2)
\nonumber \\
 {\rm E}_4 &\rightarrow&  \quad  
  \gamma_{{\rm E}_4} = 0 \, , \qquad   \qquad  \qquad \delta_{{\rm E}_4} = 16\epsilon
\nonumber \\
\Box R &\rightarrow&  \quad  
  \gamma_{\Box} = 0 \, , \qquad   \qquad  \qquad \, \, \delta_{\Box} = +24 \epsilon^2
\ea
The corresponding correction to the tadpole is 
\ba \label{cnt tadpole}
  \Delta   \langle \zeta_{\bf q}(\tau) \rangle_{\rm cnt} &=& -2Im\left\{ \int^{\tau}\!\!d\tau_1 
  \langle H_{\rm lin}\zeta_{\bf q}(\tau) \rangle \right\}
\nonumber \\
  &=& \delta({\bf q}) \times  (\delta - \gamma) \,
 q^3 \vert \zeta_q^0 \vert^2 \left\{\, \ln(-q\tau) + {\rm cte} + O(q\tau)  \right\} 
\ea
Note that although neither $\gamma_{R^2}$ nor $\delta_{R^2}$ are of order $\epsilon$,
their difference is. This will prove important in the following section.
Similarly for $G_N$. We also see that the cosmological constant has a vanishing
contribution.

In conclusion, it is a significant check that the counterterms are of the same order
in $\epsilon$ as the interactions they renormalize: 
$H_3$ induces corrections of $O(\epsilon)$ to the tadpole, which are 
renormalized by the linear terms (\ref{lin cnt}),  
and corrections $O(\epsilon^2)$ to the two point function, which are 
renormalized by the square of the Weyl tensor, while 
$H_4$ induces corrections $O(\epsilon^0)$ which are
renormalized by the counterterms (\ref{cnt zeta zeta'}) and (\ref{polynome cnt}).

\section{Renormalization of the tadpole and trace anomaly}
\label{sec:tadpole}

The renormalization of the tadpole is tantamount to the renormalization of the
energy momentum tensor since
$H_3 = - \frac{a^2}{2}\int\!\!d^3x h^{ab}T_{ab}$.
Indeed we have
\ba  \label{formal tadpole}
  \langle \zeta_{\bf q}(\tau) \rangle &=& \delta({\bf q}) \times 
  2 Im \left\{ \zeta_q^*(\tau) \int^{\tau}\!\!d\tau_1 \, a_1^2 \left( -\alpha_1 \langle T_{00} \rangle
  + \zeta_1 \delta^{ij} \langle T_{ij} \rangle \right)
  \right\}
\ea
where $2T_{00} = {\sigma'}^2 + (\partial \sigma)^2$. 
We immediately see that the approximation (\ref{modemin}) is inadequate.
If one insists on using it, one finds after integrating over time 
\ba \label{bad divergences}
  \langle \zeta_{\bf q}(\tau) \rangle &=& \delta({\bf q}) \times \vert \zeta_q^0 \vert^2 \,
  \left\{ \frac{1}{q} (1 - q^2 \tau^2) J_0 - \frac{q}{2} J_2 \right\}  
\ea
where $J_0$ and $J_2$ the $q$-independent divergent integrals
such that $T_{00} = J_0 + \frac{\tau^2}{2}J_2$.
There is no term of the list (\ref{counterterm R^2}) with the proper
$q$-dependence to absorbe these divergences. 

Of course we know how to proceed.
The renormalized value of the energy-momentum tensor of fields in curved backgrounds
is obtained from the adiabatic expansion of the modes
(the covariant methods look very different but the requirement 
of the Hadamard form expresses essentially the same idea. 
I therefore favour the more intuitive language 
of the adiabatic substraction). Since this expansion 
is close to an expansion in slow roll parameters, we understand 
that the instantaneous de Sitter approximation around horizon exit 
(\ref{modemin}) cannot do the job.

The counterterms for a minimally coupled scalar field are \cite{BirrelDavis}
\ba \label{cnt trace}
  \Delta \Gamma_{\rm cnt} 
   &=&  - \frac{\mu^{4-d}}{4-d} \, \frac{1}{5760 \pi^2} \, \left\{ 3 C^2 - {\rm E}_4
   - 12 \Box R + 5 R^2 \right\}
\ea
(The last two terms are absent for a conformally coupled scalar field.)
I make a simplification and take the renormalized value of the 
energy momentum tensor in de Sitter space 
\ba \label{exp value Tab}
  \langle T_{ab} \rangle = \beta \, \dot \rho^4 \, g_{ab}
\ea
where $\beta \neq 0$ is a constant. The trace 
$\langle  T^a_{\,\, \,a} \rangle_{\rm ren} = 4\beta \dot \rho^4 $ therefore 
does not vanish, even for conformal fields.
For minimaly coupled fields in the Allen-Follaci vacuum, $\beta = - \frac{119 {\cal N}}{960 \pi^2}$.
After substituting (\ref{exp value Tab}) and integrating over time we get
\ba \label{zero tadpole}
   \langle \zeta_{\bf q}(\tau) \rangle &=& - \beta\, \delta({\bf q}) \times 
  Im \left\{ \zeta_q^*(\tau) \int^{\tau}\!\!\frac{d\tau_1}{\tau_1^4} \, \left( 3\zeta_1 + \alpha_1  \right)
  \right\} 
\nonumber \\
   &=& 0
\ea
This is a significant result because the symmetry responsible for the 
conservation of $\zeta$ is a symmetry of the action for ${\bf q}=0$.
True, the zero mode of $\zeta$ is a gauge mode. But like a Goldstone mode
it acquires a physical relevance by the fact that it can be continuously extented 
to physical modes \cite{WeinbergAdiabmode}.
Had we any other linear combination than $3\zeta + \alpha$ in (\ref{zero tadpole}), 
we would have obtained
$\langle \zeta_{{\bf q}=0}(\tau) \rangle  \propto \ln(-q\tau)$ (as in (\ref{cnt tadpole})).
This perfect fit of coefficients bears no relation with a fine tunning. 
It is a sign of the consistency of the theory, to wit, 
the symmetry fixing (\ref{exp value Tab})
is not responsible of its own demise, as would constitute a breaking of its 
subgroup (\ref{symm q=0}).

Two other results back up this assertion.
One is that the corrections from the linear counterterms 
(\ref{list delta gamma}-\ref{cnt tadpole}) are $O(\epsilon)$, and therefore vanish 
in the de Sitter limit. This is consistent with the fact that we are neglecting 
$\zeta^3$ interactions and that linear scalar perturbations in de Sitter space 
are gauge modes.
The second result is that if we do not assume the de Sitter symmetry this 
cancellation does not happen. 
In a model of power law inflation $a(\tau) = \tau^{-1-\alpha}$
with $\alpha \geq 0$
one obtains for $q\tau \leq 1$
(see appendix \ref{app:time int red spectrum})
\ba
  \langle \zeta_{\bf q}(\tau) \rangle &=& \# \delta({\bf q}) \times q^3 (q\tau)^{\alpha} 
\ea
which is finite for a finite value of $\tau$ and $\to 0$ in the asymptotic future.
Moreover, the $\epsilon \ln(-q\tau)$ in (\ref{cnt tadpole}) 
is also replaced by $(q\tau)^{\alpha}$, which shows that the linear terms 
(\ref{lin cnt}) are of the appropriate form to renormalize the tadpole.

\section{Correction from two insertions of $H_3$}
\label{sec:three vertex}

Instead of calculating the corrections from \Eqref{H_3}, 
it is possible to make a cunning canonical transformation such 
that the scale dependent correction $q^3\ln(q/\mu)$ has a unique antecedent \cite{Weinberg}
(the corresponding transformation for the $\zeta^3$ interaction is given in \cite{Maldacena}).
I quote the result. 
Using again the notation $Q(t)=\zeta_{\bf q}(t) \zeta_{-{\bf q}}(t)$, we have \cite{Weinberg}
\ba \label{perturb}
  \Delta {\cal P} &=&
  - \int_{-\infty}^{t}\!\!dt_2  \int_{-\infty}^{t_2}\!\!dt_1\, 
  \langle \left[\widetilde H_3(t_1),\, \left[ \widetilde H_3(t_2),\, Q(t)  \right]    
 \right] \rangle \nonumber \\
  &&\quad - \, \int_{-\infty}^{t}\!\!dt_1 \,  
  \langle \left[ \left[{\cal F}_1,\, \widetilde H_3(t_1) + \frac{1}{2}\dot {\cal F}_1 \right] 
  ,\, Q(t)  \right] \rangle 
  -  \langle \left[{\cal F}(t),\, \left[ {\cal F}(t),\, Q(t)  \right]    
 \right] \rangle  + ...
\ea
The new vertices $\widetilde H_3$ and $\cal F$ are 
\ba \label{Aterm}
  \widetilde H_3 &\equiv&  a^3 \int\!\!d^3x \, 
 \left( T_{00} + T^{i}_{\,\, i}   \right) 
  \left(-\epsilon \dot \rho a^2 \partial^{-2} \dot \zeta \right)
\\
  \label{Bterm}
   {\cal F} &\equiv& a^5 \int\!\!d^3x \,  B T_{00}
\ea
One clearly sees from (\ref{perturb}) that only (\ref{Aterm}) can be responsible for 
a $q^3\ln(q/\mu)$ because the term (\ref{Bterm}) generates effective fourth order interactions
which produce polynomial corrections $q^3$.

I now make a remark that will shortly be useful: one recognizes $\widetilde H_3$ as
$-\frac{1}{2}\int\!\!d^4x \sqrt{-g} \, \delta g_{ab} T^{ab}$ 
in the longitudinal gauge, 
where the gravitational potential $\phi_l$ and the perturbation of 
the three metric $\psi_l$ are equal and given by 
\ba \label{phi_l}
   \phi_l =\psi_l = -\epsilon \dot \rho a^2 \partial^{-2} \dot \zeta
\ea
The proof of this statement can be found in appendix \ref{app:long gauge}.

I now proceed with the corrections from (\ref{Aterm}).
The finite part has already been calculated in \cite{Weinberg}, so I focus on
the renormalization. To show that the divergence can be absobed in the counterterm
$C^2$, I exploit the underlying analytic and tensor structure made obvious
by the fact that the Hamiltonian is $-\int\!\!d^3x {\cal L}_3$.

After doing the Wick contractions, the first term of (\ref{perturb}) can be written
\ba \label{1loopzeta}
  \Delta_{3} {\cal P}(q) 
  &=& -4 {\cal N} \int\!\!\frac{d^{3}p}{(2\pi)^3} \, \frac{d^3p'}{(2\pi)^3} \, 
 (2\pi)^3\delta^{(3)}\left( {\bf q} + {\bf p} + {\bf p}'  \right)\, 
 \nonumber \\
 && \qquad \qquad 
 \times\, \int_{-\infty}^{\tau}\!\!d\tau_2 V_2  \int_{-\infty}^{\tau_2}\!\!d\tau_1 V_1 \, 
 Re\left( {\cal Z}_q^{\zeta \zeta} {\cal M}_{pp'}^{\sigma}\right) 
\ea
${\cal N}$ is the number of scalar fields.
The vertex function is $V(\tau) = -\epsilon H a^6$. 
The function ${\cal Z}_q$ comes from the Wick contractions of 
$\langle \dot \zeta_1 \dot \zeta_2 \zeta_t^2 \rangle - \langle \dot \zeta_1 \zeta_t^2 \dot \zeta_2 \rangle$.
It is given by
\ba \label{Z}
  {\cal Z} = \frac{1}{q^4a_1 a_2} \, \zeta_{q}'(\tau_1)  \zeta_{q}^*(\tau) \left[ 
  \zeta_{q}'(\tau_2)  \zeta_{q}^*(\tau) - 
  \zeta_{q}(\tau)  {\zeta_{q}'}^*(\tau_2) 
  \right] 
\ea
${\cal M}$ comes from the Wick contractions of 
$ \langle \dot \sigma^2(t_1)  \dot \sigma^2(t_2) \rangle$,
\ba \label{M}
  {\cal M}^{\sigma}_{pp'}(\tau_1, \tau_2) = 
  \frac{2}{a_1^2 a_2^2} \sigma_p'(\tau_1)  {\sigma_p'}^*(\tau_2) \, 
  \sigma_{p'}'(\tau_1)  {\sigma_{p'}'}^*(\tau_2)
\ea
If in \Eqref{1loopzeta} we factor out the normalization of the metric modes we obtain 
\ba \label{compact1loop}
  \langle \zeta_{\bf q}(t)  \zeta_{-\bf q}(t)  \rangle
 &=& - 16 {\cal N} \epsilon_q^2 \vert \zeta_q^0 \vert^4 \,  
  \int_{-\infty}^{\tau}\!\!d\tau_2  \int_{-\infty}^{\tau_2}\!\!d\tau_1 \,  
  \,{\rm Re} \left\{ \left( 1-iq\tau \right)^2 \widetilde \Sigma_> 
   - \left( 1 + q^2\tau^2 \right)\widetilde \Sigma_<   \right\}
\qquad   
\ea
where 
\ba
  \tilde \Sigma_> &\equiv& e^{-iq(\tau_1 + \tau_2 - 2\tau)} \, 
  \int\!\!\frac{d^{3}p}{(2\pi)^3} \, \frac{d^3p'}{(2\pi)^3} \, 
  (2\pi)^3\delta^{(3)}\left( {\bf q} + {\bf p} + {\bf p}'  \right)\, 
  {\cal M}_{\rm red}(p,p',\tau_1,\tau_2) 
\qquad
\\
  \tilde \Sigma_< &\equiv&  e^{-iq(\tau_1 - \tau_2)} \, 
  \int\!\!\frac{d^{3}p}{(2\pi)^3} \, \frac{d^3p'}{(2\pi)^3} \, 
  (2\pi)^3\delta^{(3)}\left( {\bf q} + {\bf p} + {\bf p}'  \right)\,
  {\cal M}_{\rm red}(p,p',\tau_1,\tau_2) 
\qquad
\ea
and 
\ba \label{Mred}
  {\cal M}_{\rm red}(p,p',\tau_1,\tau_2) \equiv \left( pp' \right)^2 \, 
  \frac{e^{-i(p+p')(\tau_1 - \tau_2)}}{2p \, 2p'}
\ea
The latter is indentical to the positive Wightman function
$\langle \left( \partial_{\tau_1} \chi(\tau_1) \right)^2 \, 
 \left( \partial_{\tau_2} \chi(\tau_2) \right)^2 \rangle$
of a scalar field $\chi$ in Minkowski space in its ground state.
We notice that the factors of $a$ and $\dot \rho$ cancel.

It is a simple exercice to verify that 
the expression \Eqref{1loopzeta} is the Fourier transform of
\ba \label{1loopbis}
  \Delta G_F(\tau,\tau',{\bf x}-{\bf y}) &=& 
   i {\cal N}\, 
   \int_{-\infty}^{\tau}\!\!d\tau_2 d^3z_1 \int_{-\infty}^{\tau_2}\!\!d\tau_1 d^3z_2 \, 
   \left( G_>^{\zeta}(\tau,{\bf y};\tau_2,{\bf z}_2) -  
   G_<^{\zeta}(\tau,{\bf y};\tau_2,{\bf z}_2) \right) \, 
\nonumber \\ 
  &\times& \, 
  \left[  G_>^{\zeta}(\tau,{\bf x};\tau_1,{\bf z}_1) 
  \Sigma_>^{\sigma}(\tau_1,{\bf z}_1;\tau_2,{\bf z}_2) -
  G_<^{\zeta}(\tau,{\bf x};\tau_1,{\bf z}_1) 
  \Sigma_<^{\sigma}(\tau_1,{\bf z}_1;\tau_2,{\bf z}_2)
  \right]
\qquad
\ea
where $G_>^{\zeta}$ is  
\ba \label{G>}
  G_>^{\zeta}(\tau,\tau_1) = G_<^{\zeta}(\tau_1 , \tau) &=& 
  \int\!\!\frac{d^3q}{(2\pi)^3} \, e^{i {\bf q}({\bf x} - {\bf z})} 
  \zeta_q(t) \frac{\dot \zeta_q^*(t_1)}{q^2}
\nonumber \\
  &=& -\frac{1}{a_1^2 \dot \rho_1} \, 
  \int\!\!\frac{d^3q}{(2\pi)^3} \, e^{i {\bf q}({\bf x} - {\bf z})} 
  \vert \zeta_q^0 \vert^2 (1+ i q\tau) e^{-iq(\tau - \tau_1)}
\ea
It is the positive Wightman function of $\zeta$ with the substitution 
$\zeta_q^*(t_1) \mapsto \zeta_q^*(t_1)/q^2$ for the leg going into the loop.
The self-energies are the connected part of the fluctuation of the 
energy-momentum tensor of the scalar field
\ba \label{Sigma>}
  \Sigma_>^{\sigma}(\tau_1,\tau_2) &=& \Sigma_<^{\sigma}(\tau_2,\tau_1)
\nonumber \\
  &\equiv& 4  V(\tau_1) V(\tau_2) \, 
  \langle \dot \sigma^2(t_1)  \dot \sigma^2(t_2) \rangle_{\rm con} 
\nonumber \\
  &=& 4\epsilon^2 a_1^2  \dot \rho_1 a_2^2  \dot \rho_2 \, 
  \int\!\!\frac{d^3p}{(2\pi)^3} \, \int\!\!\frac{d^3p'}{(2\pi)^3} \, 
  e^{i({\bf p}+{\bf p'})({\bf z}_1 - {\bf z}_2)}  \, 
  {\cal M}_{\rm red}(p,p',\tau_1,\tau_2)
\qquad
\ea
where ${\cal M}_{\rm red}$ is defined at eq. \Eqref{Mred}.
The conjugation relations $G_>^{\zeta}(\tau,\tau_1) = [G_<^{\zeta}(\tau,\tau_1)]^*$ and 
similarly for $\Sigma_>^{\sigma}$ show that the r.h.s. of \Eqref{1loopbis} is real.

This game of rewriting has a purpose. We have put eq. (\ref{1loopzeta}) into the 
generic form (\ref{1loopbis}) of the one-loop correction
of the two-point function in the in-in formalism (see Appendix \ref{app:in-in}
for an outline). Moreover, we see on (\ref{Mred}), (\ref{G>}) and (\ref{Sigma>}) 
that the various factors of $a$ and $\dot \rho$ cancel before integration
and the only time dependence is through oscillating functions.
This is an important remark since it implies that, 
as far as the analytic structure is concerned, 
except for the external legs and the finite upper bounds of the time integrals, 
everything is identical to the 
calculation of the propagator of a scalar field $\varphi$ interacting with a 
scalar $\psi$ with vertex $\varphi \psi^2$ in Minkowski space.
In particular, the self-energies 
$\Sigma_>$ and $\Sigma_<$ are related to the time-ordered self-energy by
\ba
  \Sigma_F(x-y) &=& \theta(x^0 - y^0) \Sigma_>(x,y) + \theta(y^0 - x^0) \Sigma_<(x,y)
\ea
and the counter Lagrangian is obtained from the residue of the pole
(in dimensional regularization) of the time ordered version of ${\cal M}_{\rm red}$
\ba
  \Sigma_F^{\sigma}(q^2) \propto  \int\!\!\frac{d^dp}{(2\pi)^d}  
  \, \frac{d^dp'}{(2\pi)^d} \, (2\pi)^d \delta^{(d)} \left( q + p + p'  \right)
  \frac{-i}{p^2 + i\epsilon} \,  \frac{-i}{(p')^2 + i\epsilon} \, 
 \left({\bf p}^2 {\bf p'}^2 \right)^2
\qquad
\ea

In order to find the correct combination of curvature tensors we need to 
identify the tensor structure giving the term $\left({\bf p}^2 {\bf p'}^2 \right)^2$.
This is where the remark made at the begining of this section becomes useful.
The vertex $\widetilde H_3$ is nothing but 
$-\frac{1}{2}\int\!\!d^4x \sqrt{-g} \, \delta g_{ab} T^{ab}$ 
(with $h_{00} = h_{i,j=i} = \phi_l$ given by \Eqref{phi_l}).
Since the tensor structure is the same as in the Lagrangian formalism, 
the counterterms must also be the same.
We therefore conclude that the corresponding self-energy operator is
\ba
  \Sigma_F^{\mu\nu\alpha\beta}(q^2) =  \int\!\!\frac{d^dp}{(2\pi)^d}  
  \,  \frac{-i}{p^2 + i\epsilon} \,  \frac{-i}{(q+p)^2 + i\epsilon} \, 
  \eta^{\mu \rho \nu \sigma}  \eta^{\alpha \lambda \beta \kappa} \,
  p_{\rho} \left( p_\sigma + q_\sigma \right) 
  p_{\lambda} \left( p_\kappa + q_\kappa \right) 
\ea
where
\ba
  \eta^{\mu \rho \nu \sigma} = \eta^{\mu \rho} \eta^{\nu \sigma} - \frac{1}{2} 
  \eta^{\mu\nu} \eta^{\rho \sigma}
\ea
which has already been calculated in e.g. \cite{tHooftVeltman,Duff,Tomboulis,
HartleHorowitz,CamposVerdaguer}.
The counterterm is
\ba \label{cntH_3} 
  \Delta \Gamma = - \frac{1}{120 \pi^2}\, \frac{\mu^{4-d}}{(4\pi)^2} \frac{1}{4-d} 
  \, \int\!\!d^4x \,  C_{abcd}(x) C^{abcd}(x)
\ea
in addition to $R^2$ (see the comment $(2)$ in sec. \ref{sec:cnt R^2}).
To recapitulate, this conclusion was reached after noticing that 
the analytic and tensor structure of the loop are identitical in 
the Hamiltonian and Lagrangian formalism. The 
counterterm must therefore be the same.

We can check this by a direct calculation.
Notice that (\ref{H_W}) depends only on the combination $\epsilon \rho' \zeta'$ 
as $\widetilde H_3$
and is therefore adequate to renormalize its contribution.
Combining \Eqref{compact1loop} with \Eqref{Delta Weyl}, we get
\ba \label{delta_3 P}
  \Delta_{H_A}  {\cal P} + \Delta_{\rm W} {\cal P} 
   &=& - 32\left( \epsilon \vert \zeta_q^0 \vert^2 \right)^2 q^3 (1-q^2\tau^2) 
  \left\{  {\cal N} {\cal J}\left( \frac{q}{\mu} \right) - C_{\rm W} \right\}
\ea
The dimensionaly regularized integral ${\cal J}$ is \cite{Weinberg}
\ba
  {\cal J}\left( \frac{q}{\mu} \right) = - \frac{\pi}{15} \left( \frac{1}{3-d} - \ln\left( \frac{q}{\mu}  \right) \right)
\ea
which fixes the value of $C_{\rm W}$ to 
\ba
  C_{\rm W} =  \frac{\pi {\cal N}}{15} \, \frac{1}{3-d}
\ea

\section{Correction from fourth order interactions}
\label{sec:corr H_4}

I now turn to the corrections from the fourth order interactions
\ba \label{tadpolecommut}
   \Delta_4 {\cal P}(q)  &=& 
  - 2 Im \int^{t}_{-\infty(1-i\epsilon)}\!\!\!dt_1 \, \langle H_4(t_1)\zeta_{\bf q}(\tau) \zeta_{-{\bf q}}(\tau) \rangle
\ea
I first consider the effective fourth order interactions of the previous section
(the last two terms of (\ref{perturb})). They turn out to be problematic. 
I continue with the terms \Eqref{H_4}. 
I show that: 1) the contribution of \Eqref{H_4constrains} vanishes.
This means that first solving the constraints and then quantizing the 
theory introduces no correction at one loop, which is a non trivial result.
2) So do the corrections from (\ref{H_4extra}).
3) The remaining terms are renormalized by the same counterterms as the tadpole.
Contrary to the latter, a secular dependence $q^3 \ln(-q\tau)$ appears.

\subsection{Remaining terms of eq. \Eqref{perturb}}
\label{sec:contrib F}

Let us begin with the term 
\ba \label{F-A}
  \left[{\cal F},\, \widetilde H_3 \right] &\rightarrow& 
  2Re \int^\tau\!\!d\tau_1 \, \left\{ \langle {\cal F}_1 \widetilde H_3(\tau_1)Q(\tau) \rangle 
  -  \langle Q(\tau){\cal F}_1 \widetilde H_3 (\tau_1) \rangle \right\}
\nonumber \\
  && = \, 2 \frac{\epsilon}{q^2} Re\left\{  \int^\tau\!\!d\tau_1 \, \rho' \left( \omega_q \zeta_q' [\zeta_q^*(\tau)]^2 - c.c. \right)
 a^4\langle T_{00} \sigma'^2 \rangle \right\}
\ea
When not specified, the argument of the functions under the integral sign is $\tau_1$.
Since the part coming from the Wick contractions of the $\zeta$'s is purely imarginary, we only need 
to calculate the imaginary part of $\langle T_{00} \sigma'^2 \rangle$. The result is
\ba
 \left[{\cal F}, \, \widetilde H_3  \right] \rightarrow - \epsilon(3-2\epsilon) \vert \zeta_q^0 \vert^4 \, \frac{J_2(q)}{q}
\ea
with the divergent intergral 
\ba \label{J_2}
  J_2(q) = \int\!\!\widetilde{d^3p}\, \widetilde{d^3k} \, 
  (2\pi)^3 \delta({\bf q} + {\bf p} + {\bf k}) \,
  \frac{{\bf p}\cdot{\bf k} }{pk} (p+k) \sim q^4
\ea
The divergent coefficient has the correct dependence to be absorbed into 
 (\ref{polynome cnt}).

Things do not work so well for the other two interactions. 
The second term is
\ba
  \left[{\cal F}, \, \dot  {\cal F} \right] &\rightarrow&   
 Re \int^\tau\!\!d\tau_1 \, \left\{ \omega_q \left( \omega_q' + 2\rho'\omega_q \right) [\zeta_q^*(\tau)]^2 - c.c. \right\}
\, a^4\langle T_{00} T_{00} \rangle({\bf q})
\nonumber \\
  && + \, Re \int^\tau\!\!d\tau_1 \, \left\{ \omega_q^2  [\zeta_q^*(\tau)]^2 - c.c. \right\}
\, a^4\langle T_{00} T_{00}' \rangle({\bf q})
\ea
Since $\langle T_{00} T_{00} \rangle$ is real, we are left with the second term.
Let us write $Im \langle T_{00} T_{00}' \rangle({\bf q}) = K_0 + \tau_1^2 K_2$ 
with the $q$-dependent divergent integrals
\ba
  K_0 &=& \int \vert \sigma_p^0 \vert^2  \vert \sigma_k^0 \vert^2 
\left\{ 2 k^2 p^2 (k+p) + \frac{1}{2} {\bf p} \cdot {\bf k} \left( {\bf p} \cdot {\bf k} - p^2 \right) \right\}
\nonumber  \\
  K_2 &=&  \int \vert \sigma_p^0 \vert^2  \vert \sigma_k^0 \vert^2 k^3p^2 \left\{  
  {\bf p} \cdot {\bf k} \left( {\bf p} \cdot {\bf k} - p^2 \right) + p ^2 \right\}  
\ea
The measure is the same as in \Eqref{J_2}.
We are left with
\ba
 Im  \int^{\tau}\!\!d\tau_1 \, \omega_q^2 a_1^4 \left( K_0 + \tau_1^2 K_2 \right)
\ea
to calculate.
The integrals $K_0$ and $K_2$ have respective dimensions $q^2$ and $q^4$.
After the time integration, they are multiplied by $q^{-1}$, giving thus
a correction $ A\, q \vert \zeta_q^0 \vert^4$. The divergent coefficient 
$A$ cannot be renormalized by any of the counterterms.
We face the same difficulty with the last term 
\ba
  \langle \left[{\cal F}(t),\, \left[ {\cal F}(t),\, Q(t)  \right]  \right] \rangle  
  &\rightarrow& Re\left\{ \omega_q \zeta_q^* \left( \omega_q \zeta_q^* - \omega_q^* \zeta_q  \right)  \right\} 
\times a^4(\tau) J_3(q,\tau) 
\ea
The part depending on $\zeta$ is 
$q^2 \frac{a^4}{H^4} \left( \epsilon \vert \zeta_q^0\vert^2\right)^2$, and the 
part depending on matter is
\ba
 J_3(q,\tau) = \int\!\!\widetilde{d^3p_1} ... \widetilde{d^3k_2} \, \delta({\bf q} + {\bf p}_1 + {\bf k}_1)
\, \delta({\bf q} + {\bf p}_2 + {\bf k}_2)\, \langle T_{00}({\bf p}_1,{\bf k}_1) T_{00}({\bf p}_2,{\bf k}_2)  \rangle(\tau)
\ea

We stumble against the same nonrenormalizable divergencies 
as with the tadpole (\ref{bad divergences}), for which 
the approximation (\ref{solzeta}) was bearing the responsibility.
We can reasonnably assume that both problems have the same solution.
What distinguishes $[ {\cal F}, \, \widetilde H_3 ]$
from the other two is that it is 
the only one with $\widetilde H_3$.
As for (\ref{1loopzeta}), the derivative couplings 
make things inside the loop look like we were calculating 
on a Minkowski background, 
for which we do not need to invoque the adiabatic expansion.

\subsection{No gauge anomaly}

I show that $H^{\rm \alpha,B}_{4}$ does not contribute to the power spectrum 
at one loop. I begin with an elementary remark.
The terms of the form
\ba
   \left( \dot \sigma \partial_i  \sigma \right)  V^i \, , \qquad 
   \partial^{-2} \left( \dot \sigma \partial_i\sigma \right)  V^i
\ea
have a vanishing contribution in the vacuum because of rotational symmetry.
Several terms can be cast in this form after an 
integration by parts over the spatial coordinates:
\ba
 \partial_i w_m^j \, \partial_i w_g^j 
    \,\,  , \quad 
 4V\alpha_g \alpha_m 
    \,\, , \quad 
 - 4\partial^2 B \,  \partial_i B_m  \, \partial_i \zeta
    \,\,  , \quad  
  2 \left(3\zeta - \frac{\dot \zeta}{\dot \rho} \right)\partial_i\partial_j B  \, \partial_iw_m^j 
\ea 
This leaves us with three terms, all containing a $B_m$.
I now show that their contribution also vanishes in the vacuum.
Consider for instance $\partial^2 B_m  \, \partial_i B \,  \partial_i\zeta$, which gives
\ba
  \int^t_{-\infty}\!\!dt_1\, a^3(t_1) \int\!\!\widetilde{d^3p}\,\, {\bf p}^2 \, \langle B_{m}({\bf p}) \rangle
   \int\!\!\widetilde{d^3l}\widetilde{d^3l'} \,\,  {\bf l}\cdot {\bf l}'\, 
     \langle \left[ B_{{\bf l}} \zeta_{{\bf l}'} ,\,  \zeta_{{\bf q}}  \zeta_{-{\bf q}} \right] \rangle
\ea
Refering to expressions  \Eqref{alpha_m} and \Eqref{B_m}, 
the Fourier transform of $B_m$ is
\ba \label{<B_m>}
  4\dot \rho \, {\bf p}^2 \langle B_m({\bf p}) \rangle &=&
   \delta({\bf k} + {\bf k}' - {\bf p}) \langle 
   \dot \sigma_{\bf k}  \dot \sigma_{{\bf k}'} 
 - \frac{{\bf k} \cdot {\bf k}'}{a^2} \sigma_{\bf k}  \sigma_{{\bf k}'}  
  + 4 V \alpha_m({\bf p})
  \rangle
\nonumber \\
  &=& \delta({\bf p}) \left[\int\!\!\widetilde{d^3k} \,\left( \vert \dot \sigma_{k} \vert^2 
   + \frac{k^2}{a^2} \vert \sigma_{k} \vert^2\right) +  4 V \langle \alpha_m({\bf k}=0) \rangle \right] 
\ea
so that the integral $\int\!\!\widetilde{d^3p}\,\, {\bf p}^2 \, \langle B_{m}({\bf p}) \rangle$
is just the term of \Eqref{<B_m>} inside the brackets.
To see that this term vanishes identically, 
turn to the energy constraint at second order
\ba
  &&- 6\dot \zeta^2 + 12\dot \rho \,  \partial_i B  \, \partial_i \zeta + 4 \dot \zeta  \, \partial^2 B 
  + 4 \dot \rho  \, \partial^2 B_2 + \partial_i\partial_j B  \,  \partial_i\partial_j B - \partial^2 B \,  \partial^2 B
  + \dot \sigma^2 + \frac{(\partial \sigma)^2}{a^2} 
\nonumber \\
  && \qquad = -8 \alpha \frac{\partial^2 \zeta}{a^2} 
  - 8 \zeta  \, \frac{\partial^2 \zeta}{a^2} + 2\frac{(\partial \zeta)^2}{a^2} - 2 \alpha^2 V 
  - 4 \alpha_2 V
\ea
and take it's Fourier transform, followed by the limit $q \to 0$.
The result is
\ba
   4 V \langle \alpha_m({\bf q} = 0) \rangle   =  - \int\!\!\widetilde{d^3p} \, 
  \left(\vert \dot \sigma_{p} \vert^2 + \frac{p^2}{a^2} \vert \sigma_{p} \vert^2 \right)
\ea
which proves the announced result.
The contributions of $\left(3\zeta - \dot \zeta/\dot \rho \right)\partial^2 B  \, \partial^2 B_m$ 
evidently vanishes for the same reason, and so does the contribution of
$\left(3\zeta - \dot \zeta/\dot \rho \right)\partial_i\partial_j B   \, \partial_i\partial_j B_m$
because 
\ba
  \int\!\!\widetilde{d^3p}\,\, p_i p_j \, \langle B_{m}({\bf p}) \rangle = \frac{\delta_{ij}}{3}
  \int\!\!\widetilde{d^3p}\,\, {\bf p}^2 \, \langle B_{m}({\bf p}) \rangle 
\ea
In conclusion, $H^{\rm \alpha,B}_{4}$ gives a vanishing contribution at one loop.

\subsection{Remaining contributions of $H_{4}$ and trace anomaly}
\label{sec:H_4}

The terms $H_{\rm extra}$ do not contribute. 
To see this, first notice that the expectation value of the matter fields
produces a $\delta({\bf l}+{\bf l}')$ which combined with the other 
Dirac distributions of eq. (\ref{H_4extra}) yields $\delta({\rm \bf q})$.
Second, recall from the discussion below equation (\ref{def c,d,...}) 
that the coefficients $C,\tilde D,...$ vanish for the zero mode of $\zeta$. 
This implies that the whole contribution of $H_{\rm extra}$
vanishes at one loop.

The treatment of the remaining corrections is very similar to the 
renormalization of the tadpole. It pays off to write 
the remaining terms of $H_4$ exhibiting $T_{ab}$,
\ba \label{formal delta_4 P}
  \Delta_4 {\cal P} &=& - Im \left\{ \left[ \zeta_q(\tau)^* \right]^2 \,  \int^\tau\!\!d\tau_1 \, 
   3\zeta_q^2 a^2 \langle T_{00} \rangle + 2 (\zeta_q^2 - \zeta_q \alpha_q) a^2 \delta^{ij} \langle T_{ij} \rangle \right\}
\ea
After substitution of the expression (\ref{exp value Tab}) of the renormalized energy-momentum tensor
and integration over time, one finds
\ba \label{delta_4 P}
    \Delta_4 {\cal P}_{\rm ren} &=& 
   - 8 \beta \, q^3 \vert \zeta_q^0 \vert^4 \left\{ \, \ln(-q\tau) + {\rm cte} + O(-q\tau) \right\}
\ea
We note that this correction is identical to the one of 
the counterterm ${\rho'}^3 \, \zeta \zeta'$ which is the last counterterm that
had not yet found a use.

\section{Discussion}
\label{sec:discussion}

\subsection{Summary of the results}

The technical point treated in this work is the renormalization of the 
correlation function of $\zeta$ from matter loops in the Hamiltonian formalism.
The reasons for this are explained in sec. \ref{sec:explain}.
At one loop order contribute two types of terms, the first with two insertions of the 
three vertex $\zeta \sigma \sigma$ \Eqref{H_3}, the second with one insertion
of the four vertex $\zeta \zeta \sigma \sigma$ \Eqref{H_4}.

But with one exception, I checked that the divergences can be renormalized by the same 
counterterms as in the Lagrangian formalism.
I argued in section \ref{sec:contrib F} that this exception is a technical problem,
not a fundamental one, and I leave its resolution to future work. 
I showed that the tadpole can be consistently renormalized to zero.
Concerning the power spectrum, I showed that the corrections from the three
vertex are renormalized by the square of the Weyl tensor.
Concerning the corrections from fourth order interactions, I showed that 
the terms generated by fixing the gauge and inverting the relation 
between fields and their conjugate momenta 
have a vanishing contribution, and I finally showed that the 
divergences can be absorbed in the counterterms ${\rm E}_4$,
$R^2$ and $\Box R$.

We can finally stage the final result, and make contact with the introduction.
The relative correction to the power spectrum at late time is 
\ba \label{univpower}
  \frac{\Delta {\cal P}_{\zeta}}{{\cal P}_{\zeta}} =  - 
  {\cal B} \, \epsilon GH^2  \ln\left( \frac{q}{\mu}  \right) - {\cal A} GH^2  \ln\left( -q\tau  \right) + O(q^3)
\ea
where ${\cal A}$ and ${\cal B}$ are constants and $\mu$ is the arbitrary renormalization scale.
The second term comes from an asymptotic expansion, valid for $q\tau \ll 1$.
The invariance of the action under dilatation was 
identified as the symmetry principle responsible for the conservation law
of $\zeta$. We have now verified that at one loop, the trace anomaly 
can be the cause of a violation of this invariance.

The following three comments concern the ${\cal B}$-term. 
They aim mostly at showing the physical transparency of the Hamiltonian formalism.
In particular, comments \ref{sec:AB anomalies} and \ref{sec:unitarity}
are not new but make contact with well known
results obtained in the Lagrangian formalism.
The comment \ref{sec:time integrals} concerns the choice of the approximation of the modes
and its incidence on the ${\cal A}$-term responsible for the violation.
Section \ref{sec:backreaction} is an attempt to open up perspectives on the backreaction problem.

\subsection{Physical interpretation of the counterterms}
\label{sec:AB anomalies}

Two of the counterterms were found to be
\ba
  \int\!\!\sqrt{-g} \, C^{abcd}C_{abcd} 
\quad , \qquad  
  \int\!\!\sqrt{-g} \, \left( R^{abcd}R_{abcd}  - 4 R^{ab}R_{ab} + R^2  \right)
\ea
in addition to $R^2$ and $\Box R$.
They are associated with topologically distinct graphs giving analytically 
distinct corrections, $\epsilon q^3 \ln(q/\mu)$ and $q^3$ respectively.
This is in perfect agreement with what we know about the relation of these 
counterterms to the trace anomaly.
Let me briefly recall these facts.

The square of the Weyl tensor and the Euler density 
indeed have two different geometrical and physical meanings.
In the nomenclature of \cite{typeAB anomaly}, 
the former is called the type B anomaly and 
corresponds to an effective action which is not invariant under dilatations 
since it depends on the renormalization scale $\mu$
\ba \label{Gamma non local}
   \Gamma = - \frac{1}{3840\pi^2} \int\!\!d^4x \, \sqrt{-g} \, C^{abcd} \ln\left( \frac{\Box}{\mu^2} \right) C_{abcd} 
\ea
The Gauss-Bonnet term (type A anomaly) on the other hand 
is the unique integrated scalar density (quadratic in the curvature) 
which is invariant by dilatation.
We understand in those terms that $C^2$ 
renormalizes the correction from (\ref{Aterm}), since it is the only origin of the
$\ln(q/\mu)$, while the Gauss-Bonnet term renormalizes 
the polynomial corrections from the fourth order interactions.

\subsection{Further role of the energy-momentum tensor}
\label{sec:role T_ab}

The ${\cal B}$-type corrections $\propto q^3 \ln(q/\mu)$  
from conformal fields 
(scalar, Dirac fermions and gauge vector fields) 
have been calculated in \cite{Weinberg2} and \cite{Weinbergstudent}.
The stricking result is that they 
only differ from each other and from the case of a 
minimally coupled scalar by the value of the constant ${\cal B}$.
Moreover, this constant has a universal sign, to wit it is positive
for every type of matter field.
I explain in turn in this section and the following one the physical mechanism 
behind these two results.

The universal form of the ${\cal B}$-corrections 
is really non trivial because of the different time behaviours 
of the mode functions of minimaly and conformaly coupled fields.
Conformal modes \Eqref{modeconf} redshift, i.e. their amplitude $\propto a^{-1}$,
while minimaly coupled fields \Eqref{modemin} are parametrically amplified and asymptote to a constant 
(this is how the primordial spectra are generated by inflation).
Naively, one would therefore expect
loop corrections from the latter to be enhanced by the amplification factor
$(aH/q)^2$.

This is not so because, and only because, the metric perturbations couple to 
the energy momentum tensor. This is revealed by an inspection of $\widetilde H_3$,
\ba \label{vertexA}
  \widetilde H_3 =   a^4 \int \!\!d\tau \phi_l  \left( T^{00} + T^{i}_{\,\,i}  \right)
  \rightarrow  \epsilon a^4 \zeta_q^0 e^{-iq\tau}
  \left( T^{00} +  T^{i}_{\,\,i}   \right)(\tau_1)
\ea
I used the expression \Eqref{solzeta} 
and I implicitly made the Wick contractions of $\zeta$ with the external leg
carrying a wavenumber $q$.

Let us first consider conformal fields, and let us take it to be a scalar for the illustration. 
Using \Eqref{EOMxi}, we have
\ba \label{Tconf}
  \left(T^{00} +  T^{i}_{\,\,i}\right) (p,\tau)  &=&
 \vert \dot \sigma_{\bf p} + \dot \rho \sigma_{\bf p} \vert^2 + \frac{p^2}{3a^2 } 
  \vert \sigma_{\bf p}  \vert^2 
\ea
where $p$ is the momentum circulating in the loop.
Using the solution \Eqref{modeconf} of the mode equation,
we see that the combination 
\ba \label{derivmodeconf}
  \dot \sigma_{\bf p} + H \sigma_{\bf p} = i \frac{p}{a} \sigma_{\bf p}(t)
  =  i \frac{p}{a^2} \sigma_{p}^0 e^{-ip\tau_1}
\ea
scales like the gradient term in \Eqref{Tconf}, so that
\ba
  \left( T^{00} +  T^{i}_{\,\,i}  \right)({\bf p}) = \frac{4}{3} \frac{p}{a^4} \sigma_p^0
\ea
The factors of $a$ in \Eqref{vertexA} therefore cancel.
The only remaining dependence on the background is through the slow-roll parameter,
and the only dependence on $\tau_1$ and $\tau_2$ comes from the Wick contraction of
$\dot \sigma_{\bf p}(\tau_1)$ from a first vertex with $\dot \sigma_{\bf p'}(\tau_2)$
from the second vertex, producing an additional term $e^{-i(p+p')(\tau_1 -\tau_2)}$,
see Eqs. (\ref{compact1loop}-\ref{Mred}).
Hence the $\tau_1$ and $\tau_2$ appear only in phases, making the time integrals finite
for $\tau \to 0$. This is Weinberg's theorem in action (see section \ref{sec:explain}),
and this is what is expected of conformal fields.

Let us now turn to the case of a minimaly coupled scalar field. Then,
\ba \label{Tmin}
  \left(T^{00} + T^{i}_{\,\,i}\right) (p,\tau)  &=& 
  2 \vert \dot \sigma_{\bf p} \vert^2 
\ea
This quadratic form is very different from \Eqref{Tconf} but we see that
the solution of the mode equation gives,
\ba
   \dot \sigma_{\bf p}(t_1) = - \frac{p^2}{a^2 \dot \rho} \sigma_p^0 e^{-ip\tau_1}
\ea
and comparing with \Eqref{modeconf} and \Eqref{derivmodeconf}, 
we find that the factor $a\dot \rho/p$  of the amplitude $\sigma_p^0$ 
that is gained by the amplification 
is exactly compensated by an extra 
factor $p/a\dot \rho$ from the time derivative.
$\widetilde H_3$ with minimaly and conformaly coupled fields are therefore 
of the same order
\ba
 \widetilde H_3[\xi = 0] \sim \widetilde H_3[\xi = \frac{1}{6}] 
  &\sim& \epsilon \,  p \, \zeta_q^0 \, e^{-iq\tau}
\ea
They only differ throught their functional dependence on $\bf p$ and ${\bf p}'$.

For this to happen, we see that having derivative couplings is necessary but not 
sufficient, as illustrated by the
difference between \Eqref{Tconf} and \Eqref{Tmin}.
The coupling of the metric to the energy-momentum
tensor is instrumental.

\subsection{Unitarity}
\label{sec:unitarity}

From the expression (\ref{Gamma non local}) of the non local part of 
the effective action, we immediately obtain its imaginary part 
\ba
  Im \left( \Gamma \right) = \frac{1}{3840 \pi} \int\!\!\frac{d^4 q}{(2\pi)^4} \, 
  \theta(-q^2) \vert C_{abcd}(q) \vert^2
\ea
since $\ln(-q^2/\mu^2) = \ln(\vert q^2/\mu^2 \vert ) - i\pi \theta(-q^2)$.
Using the argument of \cite{Duff,Latore} based on the spectral representation and 
that I do not reproduce, we can now explain 
the fact that the constant ${\cal B}$ in \Eqref{univpower} is always positive, independently of 
the spin and the coupling: its positivity means the positivity of the 
spectral density. In other words, ${\cal B} > 0$ by unitarity.

\subsection{A comment on the de Sitter approximation}
\label{sec:time integrals}

I now discuss the de Sitter approximation employed in the previous calculations.
It is used in two instances.
One is the renormalized value of the energy momentum tensor (\ref{exp value Tab}),
the other is the approximation for the modes (\ref{solzeta}).
Jointly, they are responsible for the secular terms $\ln(-q\tau)$
in equation (\ref{delta_4 P}).

As explained in section \ref{sec:tadpole}, 
the approximation (\ref{solzeta}) is inappropriate to renormalize 
the energy-momentum tensor. It is necessary for the renormalization 
of composite operators, which are by definition
a product of fields in the coincidence limit, to use an adiabatic expansion.
By contrast (\ref{solzeta}) is only valid around the time of horizon exit and does not
capture the correct ultraviolet structure of the vacuum. 
So why is it that we seemingly did not need to use the adiabatic expansion to 
the renormalization the corrections from $\widetilde H_3$ in section \ref{sec:three vertex}, 
and can we trust the result (\ref{delta_3 P}) ? I think that we can, and 
I attribute the success of the approximation (\ref{solzeta}) in that case to 
the fact that $\widetilde H_3$ describes the conformal sector of the theory.
I mean this in the sense that the counterterm is the conformaly invariant density
$\sqrt{-g} C^2$, that $\widetilde H_3$ is identical for 
both minimaly and conformal fields (see comment \ref{sec:role T_ab}), 
and that contrary to ${\cal F}$ and 
$H_4$, it is suppressed by a factor of $\epsilon$ which measures 
the deviation of the background from the conformaly invariant de Sitter space.
These unique properties of $\widetilde H_3$ are what made a direct 
identification of the counterterm possible, 
because inside the loop, everthing looks like in Minkowski space.

I also point to another limitation of (\ref{solzeta}), at low wavenumbers this time.  
I calculated the late time expression of (\ref{delta_4 P}) in a model of power law inflation. 
The only modification with respect to the de Sitter approximation is the
expression of the modes, because the renormalized value of the energy momentum tensor
still scales like $1/\tau^4$ as in the de Sitter limit (this is because 
in both case the scale factor is a monomial in $1/\tau$).
The details are presented in appendix \ref{app:time int red spectrum}, and I quote the result
of the late time asymptotic expansion, e.g.
\ba \label{delta_4 P power law}
 \Delta_4 {\cal P} \sim   q^3 \, \vert \zeta_q^0 \vert^4 \, \frac{1}{\epsilon} 
 \left( \frac{q}{a \dot \rho} \right)^{2 \epsilon}
\ea
It is enhanced compared to the classical correction 
$(q/a\dot \rho)^2$ (on the r.h.s. of (\ref{symm q=0})), but it becomes negligeable after 
$O(\epsilon^{-1})$ efolds following the horizon crossing of the mode.
So in that case we may conclude that the correction is finite.

The outcome of this superficial analysis casts some doubts on the 
reality of the secular term (\ref{delta_4 P}), which now appears 
to be an artifact of the approximation (\ref{solzeta}).
The asymptotic decay of $\Delta_4 {\cal P}$ in (\ref{delta_4 P power law}) is the consequence 
of having a red spectrum, 
i.e. ${\cal P}(q) \sim q^{n_S - 1}$ with a spectral index $n_S < 1$.
I have two reasons for favouring this case.
One is that it is the scenario singled out by observations \cite{WMAP5}.
The other reason is more speculative: 
because a factor of $q$ always appears with a factor $1/a$, 
a spectrum with a blue tilt over a large number of decades 
produces corrections scaling like some positive power of $a(t)$.
In that case we cannot take the limit $\tau \to 0$ at any stage of our calculations.
Although the connection between the power spectrum of $\zeta$ and backreaction
is not clear (see further comments in section \ref{sec:backreaction}), 
one could presume that this behaviour is a sign of strong backreaction.
If one accepts this premise, one of two scenarios can happen. 
One is that this backreaction tends to increase the blue tilt even more.
In that unfortunate case the game is over until 
we can solve the full nonlinear problem.
The other alternative is that the backreaction tends to soften the blue tilt, 
perhaps enough to turn it into a red one. In this happy turn of event 
we can use in first approximation a red spectrum over the whole range of $q$.
Doing so, we certainly make a quantitative error on the 
amplitude of the correction, but not a qualitative one.

\subsection{Perspectives on the backreaction problem}
\label{sec:backreaction}

This brings me to my last comment.
As we learn to better handel the calculations in the Hamiltonian formalism,
we can contemplate tackling the backreaction problem with these tools.
One of the foremost advantages of the formalism is its physical transparency, 
since we work directly with the physical degrees of freedom ($\zeta$ and
the two helicities of the graviton) and use a physical clock
(the inflaton field is equal to its classical value). 
If need was, I hope to have convinced
the reader of this with the first three comments.

The theoretical obstacle to the backreaction problem is the identification
of the relevant observables to quantify the effect. 
Most of the efforts have been invested in the calculation 
of the impact of non linearities
on some measure of the local expansion rate, 
with bewildering difficulties to desentangle
the physical effect from a gauge artefact 
\footnote{\label{Backreaction} If we limit ourselves to the topic ofthe backreaction 
during inflation, a selected list of references starts with the works of
Tsamis and Woodard \cite{TsamisWoodard}    
and of Mukhanov $\&$ {\it al.} \cite{Mukhanov},
respectively on the backreaction of gravitons and of superhorizon scalar perturbations.
Their methodology was seriously criticized by Unruh \cite{Unruh critique}.
Subsequent works taking these attacks into careful consideration are 
those of Abramo and Woodard \cite{AbramoWoodard},
Geshnizjani and Brandenberger \cite{GB}, and 
Finelli $\&$ {\it al.} \cite{Finelli}. 
Although these groups consider different observables, their results 
provide evidence that, as Unruh anticipated, the backreaction of 
fluctuations on the homogeneous mode cannot be observed by local observers
who should interpret superhorizon adiabatic perturbations as a scaling of the 
coordinates, provided that the adiabatic mode decouples, i.e.
provided that the non adiabatic pressure is 
negligible, so that the inflaton is the only physical clock at their disposal.
This is also the intuition at the basis of (\ref{ratezeta}-\ref{symm q=0}).
It implies concurrently that in single field inflation, both backreaction and 
a violation of the conservation law of $\zeta$
can only be sourced by derivative couplings, in agreement with 
eq. (\ref{violation by trace}).}.

The expansion rate is but on facet of gravity, 
and there is bound to be something to be learned from 
observables sensitive more specifically to other effects, 
for instance to the decelaration parameter or to tidal forces.
I invite the reader interested in these issues to read 
the lucid discussion of Tsamis and Woodard on this point \cite{TSproposal}.
I have little to add to this, but for the following two remarks.
First, the tilt (including the running) of the power spectrum of scalar 
and tensor perturbations is an interesting quantity.
A tilt indeed measures the variation of the amplitude of the modes at 
horizon exit in response to 
a variation of the Hubble rate, via the slow roll parameters.
In other words, it measures an acceleration. 
The scale dependent corrections in (\ref{univpower})
can indeed be interpreted as a correction to the spectral index
\ba
  \Delta n_S = \frac{d \ln(q^3 {\cal P})}{d \ln q} = - {\cal B} \epsilon GH^2 
\ea
and it is interesting that unitary, i.e. ${\cal B} > 0$, implies that the backreaction
from matter fields renders the spectrum more red 
\footnote{The corrections to the slow roll 
parameters from the fluctuations of the inflaton have been 
calculated by three different groups \cite{Boyanovski,Sloth,Prokopec}. 
They each make different simplifying assumptions, employ different 
methods of calculation, and obtain different results.
In particular, one source of discrepancy can be traced back 
to their use of different gauges. Contrary to (\ref{deffoliation}), 
these gauges do not use the inflaton
as a clock, which makes difficult to assess the 
reality of the effect to local observers.}.

The second observation is that it seems reasonably feasable 
to calculate a quantity such as the 
expectation value of the squared Weyl tensor. The latter is a sum
of $n$-point functions of $\zeta$ and gravitons $\gamma$,
\ba \label{C^2 backreact}
  \langle C^{abcd}C_{abcd} \rangle 
  = 32 a^{-4} (\epsilon \rho')^2 \langle \zeta'(\tau,{\bf x})  \zeta'(\tau ,{\bf y}) \rangle 
  + O(\gamma_{ij}^2, \zeta^3)
\ea
This observable has direct physical significance since it is a measure
of the curvature. Moreover
it is a scalar which vanishes on a flat Fiedmann Robertson Walker background, 
meaning that it is a gauge invariant measure of the deviation from this background 
\footnote{Note that the contribution of the scalar perturbations is $O(\epsilon)$ 
(because of the normalization of the modes (\ref{solzeta}) contains a factor $\epsilon^{-1}$), 
so that the leading contribution comes from gravitons.
This contrasts with the intriguing results of Losic and Unruh \cite{UnruhLosic}, to wit, 
certain gauge invariant combinations of the scalar 
perturbations are dominated by second order fluctuations.
Clarifying such discrepancies should prove to be fruitful.}.

There are several technical difficulties
that need to be resolved before undertaking this calculation.
We have already been made aware of one of them, namely 
the approximation used for the modes, both in the infrared and ultraviolet.
This is nothing new however.
If we go as far as two loops, as is necessary in de Sitter \cite{TsamisWoodard},
we will have to learn how to handle overlapping divergences 
in the Hamiltonian formalism.

\acknowledgments
I am grateful to J.C. Niemeyer for his encouragements.
At an early stage I greatly profited from conversations with R. Brout and R. Parentani.
Most of this work was done while I was residing at the 
Lehrstuhl f\"{u}r Astronomie, Universit\"{a}t W\"{u}rzburg, Germany, during 
which period it was
supported by the Alfried Krupp von Bohlen und Halbach Foundation.

\begin{appendix}

\section{Change to the longitudinal gauge}
\label{app:long gauge}

All the quantities in this section are first order. 
Scalar pertubations are therefore decoupled from vector and tensor modes, in any gauge.
Consider the following parametrization of the metric and inflaton
\ba
  &&ds^2 = a^2(\tau) \left\{ - \left( 1 + 2\phi  \right)d\tau^2 + 
  2 \partial_i \omega d\tau dx^i + \left[ \left( 1 - 2\psi  \right)\delta_{ij}  + 
  2 \partial_i \partial_j E \right]dx^i dx^j
  \right\}
\nonumber \\
    &&\varphi(t,{\bf x}) = \bar \varphi(t) + \delta \varphi(t,{\bf x})
\ea
In a first order coordinate transformation parametrized by
\ba
  \tau \mapsto \tilde \tau = \tau + \xi^0 \, , 
 \qquad
  x^i \mapsto \tilde x^i =  x^i + \partial_i \xi
\ea
the five scalar perturbations defined above vary as
\ba
   &&\tilde \phi = \phi - \frac{a'}{a} \xi^0 - \xi^{0\, '} 
\, ,\quad 
   \tilde \psi = \psi + \frac{a'}{a} \xi^0  
\, , \quad
   \tilde \omega = \omega + \xi^0 - \xi' 
\, , \quad
   \tilde E = E - \xi 
\nonumber \\
  &&\widetilde{\delta \varphi} = \delta \varphi -  \bar \varphi(t) \xi^0
\ea
By definition, in the comoving and longitudinal gauges we have respectively 
\ba
  &&E_c = 0 \, , \qquad \delta \varphi_c(t,{\bf x}) = 0
\nonumber \\  
  &&E_l = 0 \, , \qquad B_l = 0 \, .
\ea
The passage from the comoving gauge to the longitudinal gauge is
therefore realized by the coordinate transformation
\ba
  \tau_c &\mapsto& \tau_l = \tau_c - \omega_c \, , \qquad x_c^i \mapsto x_l^i =  x_c^i 
\ea
i.e. $\xi^0 = -\omega_c$ and $\xi = 0$.
With $\phi_c = \frac{\zeta'}{\rho'}$, $\psi_c = -\zeta$,
$\omega_c = - \frac{\zeta'}{\rho'} + \epsilon \, \partial^{-2} \zeta'$, and the linear
equation \Eqref{EOMzeta}, one obtains
\ba
  \phi_l = \psi_l = - \epsilon \rho' \partial^{-2} \zeta'
 = - \epsilon \dot \rho a^2 \partial^{-2} \dot \zeta
\ea 
The l.h.s. of this equality is recognized as the term
multiplying $2 {\sigma'}^2 = \delta^{ab}T_{ab}$ 
(for a minimaly coupled field) in \Eqref{Aterm}.
In words, the vertex $\widetilde H_3$ is the interaction between $\sigma$ and 
the metric in the longitudinal gauge.
This can also be shown explicitely.
In a coordinate transform $x^a \mapsto x^a + \xi^a$, the action varies by 
$\delta_\xi S = - \int\!\!d^4x \sqrt{-g} \, \xi^a \nabla_b T^b_{\,\, a}$,
which induces the following correction of the $\zeta \sigma \sigma$-vertex,
\footnote{The passage to the longitudinal gauge generates of course other terms that I have not
discussed here: corrections to the
pure gravity vertices, as well as interactions with the inflaton perturbations.} 
\ba
   S_l[\zeta \sigma \sigma] &=& -\int\!\!d^4x \sqrt{-g} 
  \left\{ \frac{1}{2} \delta g_{ab} T^{ab} + \xi^a \nabla_b T^b_{\,\, a} \right\}
\nonumber \\
  &=& \int\!\!d^4x \sqrt{-g} \left\{ \frac{\dot \zeta}{\dot \rho} T^{00} - 2\partial_i \omega T^{0i} - 2\zeta T^{i}_{\,\, i}
  - \omega\left( \partial_t T^{00} + 3\dot \rho T^{00} + \partial_i T^{0i} + \dot \rho T^{i}_{\,\, i}  \right)  \right\}
\ea
Integrating by parts the second, fourth and fifth term and using the linear equation \Eqref{EOMzeta},
one finds
\ba
  S_l[\zeta \sigma \sigma] &=& \int\!\!d^4x \sqrt{-g} \phi_l \left(  T^{00} + T^{i}_{\,\, i}  \right) 
  - \int\!\!d^4x \, \partial_t\left( a^3 T^{00} \omega  \right) - \int\!\!d^4x \, \partial_i\left( a^3 T^{0i} \omega  \right) 
\nonumber \\
   &=& - \int\!\!dt \, \left(  \widetilde H_3  + \partial_t {\cal F} \right) - \int\!\!dt \int\!\!d^3x\, \partial_i\left( a^3 T^{0i} \omega  \right) 
\ea
The last term is a spatial gradient and therefore does not contribute.
The term ${\cal F}$ is now clearly identified with 
the generator of the (first order) canonical transformation 
relating the longitudinal gauge to \Eqref{defgauge}.
Indeed, the Einstein Hilbert Lagrangian density changes by a total
derivative in a gauge transformation.
So does the Lagrangian of any system in a canonical transformation.
So a gauge transformation can be seen as a canonical transformation.

\section{Expressions of the geometric counterterms}
\label{app:curvature}

\subsection{Weyl tensor}

The first counterterm is the square of the Weyl tensor. 
Since the latter vanishes for a flat Friedmann Robertson Walker space-time, we only need to
calculate it at first order.
The components of the Riemann tensor are 
\sba
  R^{0}_{\,\, i0j} &=& \left\{ \rho''(1+2\zeta) + \epsilon \rho' \zeta'  \right\} \delta_{ij} 
  +  \partial_i \partial_j\left( \epsilon \rho'\partial^{-2} \zeta'  \right)
\\
  R^{0}_{\,\, ijk} &=& O(\zeta^2)
\\
  R^{l}_{\,\, ikj} &=&  ({\rho'})^2(1+2\zeta)  \left( \delta^l_{\,\,k} \delta_{ij} -   \delta^l_{\,\,j} \delta_{ik}\right) 
 \nonumber \\
  && +\, \left[ \delta_{jl} \partial_i\partial_k + \delta_{ik} \partial_j\partial_l
  -  \delta_{lk} \partial_i\partial_j  -  \delta_{ij} \partial_k\partial_l
 \right]\left(\epsilon \rho'  \partial^{-2} \zeta' \right)
\sea
from which we obtain the components of the Ricci tensor and the Ricci scalar respectively
\sba
   R_{00} &=& -3\rho'' + (2\epsilon - 6) \rho' \zeta'  
\\
  R_{0 i} &=&  \bar {\cal R} \partial_i \omega = \left[\rho'' + 2(\rho')^2\right] \, \partial_i \omega 
\\
  R_{ij} &=& \bar {\cal R}\left( 1+2\zeta \right) \delta_{ij}
\\
  a^2 R &=& 6 \frac{a''}{a} + 4\epsilon \rho' \zeta'
\sea
The components of the Weyl tensor in four dimensions are
\ba \label{def Weyl}
  C^{a}_{\,\, bcd} = R^{a}_{\,\, bcd} - \frac{1}{2} \left( \delta^{a}_{\,\, c} \, R_{bd} - 
  \delta^{a}_{\,\, d} \, R_{bc} - g_{bc}\,  R^{a}_{\,\,d} + g_{bd} \, R^{a}_{\,\,c} \right) + 
  \frac{R}{6} \left( \delta^{a}_{\,\, c} \, g_{bd} - \delta^{a}_{\,\, d} \, g_{bc} \right)
\ea
and I find
\sba \label{Weyl}
  C^{0}_{\,\, i0j} &=& \epsilon \rho' D_{ij}\left( \partial^{-2} \zeta' \right) 
\\
  C^{0}_{\,\, ijk} &=& O(\zeta^2)
\\
  C^{i}_{\,\, jkl} &=&  \left[ 
  \delta_{jl} \, D_{ik} + \delta_{ik} \, D_{jl} - \delta_{jk} \, D_{il} - \delta_{il} \, D_{jk}  \right]\, 
  \left( \epsilon \rho' \partial^{-2} \zeta' \right) 
\sea
I introduced the notation
\ba
  D_{ij} = \partial_i \partial_j - \delta_{ij} \partial^2
\ea
One can verify that these components have the correct symmetry and 
that the identity $ C^{0}_{\,\, i0j} = - C^{k}_{\,\, ikj}$ implied by the vanishing of
the trace is satisfied.
Note that eqs. \Eqref{Weyl} depend only on the quantity (see appendix \ref{app:long gauge}) 
\ba
  \phi_l = -\epsilon \rho'\partial^{-2} \zeta'  
\ea
and have therefore the desired form to serve as counterterm for the vertex $\widetilde H_3$.
Finally, the squared Weyl tensor is 
\ba
  \sqrt{-g} C^2 &=& Ne^{3\zeta} \left\{ 4 \left( C^{0}_{\,\, i0j} \right)^2 - 4 \left( C^{0}_{\,\, ijk} \right)^2 
  + \left( C^{i}_{\,\, jkl} \right)^2 \right\}
\nonumber \\
  &=& 8(\epsilon \rho')^2  \left[ 
  \partial_i \partial_j \partial^{-2} \zeta' \, \partial_i \partial_j \partial^{-2} \zeta' 
  + 3(\zeta')^2 \right] + O(\zeta^3)
\ea
or in the momentum representation
\ba
  S_{W} &=& \frac{C_{W}}{2} \int\!\!d^4x \sqrt{-g} \,  C^2
=   \frac{C_{W}}{2}\int\!\!d\tau \, 32(\epsilon \rho')^2  
 \int\!\!\frac{d^3 p}{(2\pi)^3} \, \zeta'_{\bf p} \zeta'_{-{\bf p}}
\ea
with $C_W$ a constant.

\subsection{Second order expressions of the Ricci tensor and Ricci scalar}

For the other counterterms we need the expressions of the Ricci tensor at second order.
I give for reference the expressions of the connection coefficients 
$\Gamma^{a}_{bc} = \frac{1}{2} g^{ad} \left[ \partial_b g_{cd} + \partial_c g_{bd} - \partial_d g_{bc} \right]$:
\sba
  \Gamma^{0}_{00} &=& \rho' + \alpha' + \alpha_2' - \alpha \alpha' + \partial_i \alpha \partial_i \omega + \rho' (\partial \omega)^2
\\
  \Gamma^{i}_{00} &=& \partial_i\left( \alpha + \omega ' + \rho' \omega  + \alpha_2 + \omega_2' +\rho' \omega_2\right) 
  + w_i' + \rho' w_i  
\nonumber \\ 
  && \, + \, (2\zeta' - \alpha') \partial_i \omega + (\alpha - 2\zeta) \partial_i \alpha 
  - \partial_i \partial_j \omega \partial_j \omega 
\\
  \Gamma^{0}_{0i} &=& \partial_i\left( \alpha + \omega ' +  \alpha_2 + \omega_2' \right) + \rho' w_i 
  -\,  \zeta'\partial_i \omega - \alpha \partial_i \alpha - \partial_i \partial_j \omega \partial_j \omega  
\\
  \Gamma^{0}_{ij} &=& \delta_{ij} \left\{\rho' - \zeta' + 2\rho' \zeta - 2\rho' \alpha_2 + 2\rho' \zeta^2 
  + \rho' \alpha^2 - 2\zeta \zeta' - \partial_i \zeta \partial_i \omega  \right\} 
 \nonumber \\
  && \, - \, \partial_i\partial_j\left(\omega + \omega_2 \right) - \frac{1}{2} \left( \partial_i w_j + \partial_j w_i  \right) 
  + 2\left( \alpha - \zeta \right) \partial_i\partial_j\omega 
\\
  \Gamma^{i}_{0j} &=& \left( \rho' + \zeta' \right) \delta_{ij} - \partial_i \omega \partial_j \alpha 
  - \rho'\partial_i \omega \partial_j \omega + \frac{1}{2} \left( \partial_j w_i - \partial_i w_j  \right) 
\\
  \Gamma^{k}_{ij} &=& \partial_i \zeta \, \delta^{k}_{\, j} + \partial_j \zeta \, \delta^{k}_{\, i} - 
 \delta_{ij} \left\{ \partial_k \left( \zeta  + \rho'\omega + \rho' \omega_2 \right) 
 + \rho' w_k + (2\rho' \zeta - \zeta') \partial_k\zeta \right\}
\sea
The second order expression of the Ricci tensor are:
\ba
  \delta_2 R_{00} &=& 3 \rho' \alpha_2' 
  + \partial^2\left[ \alpha_2 + \omega_2' + \rho' \omega_2 - (\partial \omega)^2 \right]
  + \bar {\cal R} (\partial \omega)^2 -3(\zeta')^2
\nonumber \\ 
  && + \, 3 \partial_i \zeta' \partial_i \omega 
  + 3 \rho' \partial_i \zeta \partial_i \omega + 3 \partial_i \zeta \partial_i \omega' + \partial_i \zeta \partial_i \alpha
\nonumber \\ 
  && 
 + \, \left(2\zeta' - \alpha' \right) \partial^2 \omega 
 + \left(\alpha - 2\zeta \right)\partial^2 \alpha 
\ea
I separate $R_{ij}$ into diagonal and non-diagonal parts according to 
\ba
  R_{ij} = R^{\, \rm d} \, \delta_{ij} + R_{ij}^{\,\rm nd} 
\ea
with respectively
\sba
  R^{\, \rm d} &=& e^{2\zeta}\bar {\cal R} \left[ 1 + \alpha^2 - 2 \alpha_2 \right] 
 - \rho' \alpha_2' - \rho' \partial^2 \omega_2 - 3(\zeta')^2 - (\partial \zeta)^2
\nonumber \\
  && + \, 2\epsilon \rho' \zeta \zeta' - \partial_i \zeta \partial_i \alpha  
  + \left( \zeta' - 2\rho'\zeta\right) \partial^2 \omega
 - 5 \rho' \partial_i \zeta \partial_i \omega - \left[ \partial_i \zeta \partial_i \omega \right]'
\\
   R_{ij}^{\, \rm nd} &=& - \, \partial_i\partial_j \left[\alpha_2 + \omega_2' + 2\rho' \omega_2   \right]
  -\frac{1}{2} \left( \partial_i w_j' + \partial_j w_i'  \right) - \rho' \left( \partial_i w_j + \partial_j w_i  \right) 
\nonumber \\
  && +\,  \alpha \partial_i\partial_j \alpha + \left(\zeta' + \alpha' - 4\rho'\zeta \right)\partial_i\partial_j \omega
 + 2\left( \alpha -\zeta\right)\partial_i\partial_j \omega'
\nonumber \\
  && + \, \partial_i \zeta \partial_j \alpha +  \partial_j \zeta \partial_i \alpha + \partial_i \zeta \partial_j \zeta
\sea
We do not need the second order expression of $R_{0i}$ because it is a first order quantity
which multiplies other first order quantities: 
to calculate the Ricci scalar $g^{ab}R_{ab}$, it is multiplied by $g^{0i} \propto \partial_i \omega$,
and to calculate $R^{ab}R_{ab}$ it is multiplied by $R^{0i}$. 
The second order part of the Ricci scalar is
\ba
  a^2\delta_2 R &=& -12 \frac{a''}{a}\alpha_2 - 6 \rho' \alpha_2' - 2 \partial^2\left[\alpha_2 + \omega_2' + 3 \rho' \omega_2  
 - \frac{1}{2}(\partial \omega)^2 \right]
  - 8\epsilon(\zeta')^2 + 6\epsilon \rho' \zeta \zeta' - 2(\partial \zeta)^2
\nonumber \\
  && -\, 6 \left[ \partial_i \zeta \partial_i \omega \right]' - 18 \rho' \partial_i \zeta \partial_i \omega 
  - 2 \partial_i \zeta \partial_i \alpha
\nonumber \\
  && +\, 2 \zeta \partial^2 \alpha +  2 \left(\zeta' + \alpha' - 5\rho' \zeta  \right) \partial^2 \omega 
  + 2 \left(\alpha - \zeta  \right) \partial^2 \omega' 
\ea

\subsection{The remaining counterterms}

Using the solutions of the constraints and the linear mode equations to
simplify quadratic terms containing a $\zeta''$ or $\sigma''$, and dropping spatial 
divergences, I finally get
\ba \label{cnt R^2}
    S_{R^2} &=& \frac{C_{\rm R^2}}{2} \int\!\!d^4x \sqrt{-g} R^2 
 \\
   &=& \frac{C_{\rm R^2}}{2}\int\!\! \tilde R^2 \left( 1 + 3\zeta + \alpha + \frac{9}{2} \zeta^2 + 3\zeta \alpha  \right)
  + 8 \tilde R \epsilon \rho' \zeta'
  + \tilde R\left(  (2+\epsilon) T_{00} + \frac{T_{00}'}{\rho'}\right)
\nonumber \\
  && \qquad \,\,\, + \, 4\tilde R \left(\frac{1}{{\rho'}^2} (\partial^2 \zeta)^2 -  \alpha \partial^2 \alpha 
    + (1 + 5\epsilon + \epsilon_2) \, \alpha \partial^2 \zeta 
    - (3+2\epsilon) \, \zeta \partial^2 \zeta
  \right)
\nonumber \\
   && \qquad  \,\,\, + \, 36 {\rho'}^3 \zeta \zeta' 
   + 624 \epsilon (\rho' \zeta')^2 - 296 (\epsilon \rho' \zeta')^2 
\nonumber 
\ea
I used the shorthand notation $\tilde R = 6 \frac{a''}{a}$, 
and $2 T_{00} = (\sigma')^2 + (\partial \sigma)^2$.
I neglected terms of order higher than two in the slow roll parameters 
$\epsilon = 1- \frac{\rho''}{(\rho')^2}$
and $\epsilon_2 = \frac{\epsilon'}{\rho'\epsilon}$.

I give the expressions of the other counterterms only at first order in $\zeta$.
Several combinations are interesting.
The integrated Euler density reads
\ba
  \int\!\!d^4x \, \sqrt{-g} \, {\rm E}_4  &=&  \int\!\!d^4x \, \sqrt{-g} \,\left\{R^{abcd}R_{abcd}  - 4 R^{ab}R_{ab} + R^2 \right\} 
\nonumber \\
  &=&  8\int\!\!d^4x \,N e^{3\zeta} \, \left\{ 3 \rho''{\rho'}^2  + 4 {\rho'}^{\, 3} \epsilon \zeta' + O(\zeta^2)\right\}
\nonumber \\
  &=&  8\int\!\!d^4x  \left\{ -\partial_\tau( {\rho'}^{\, 3})  +  {\rho'}^{\, 3} \epsilon \zeta' + O(\zeta^2)\right\}
\ea
and I used $\rho'' = {\rho'}^2 (1-\epsilon)$ which follows from the Friedmann equations.
The linear term coming from the measure has been eliminated by an integration
by parts. The remaing term is proportional to $\epsilon \zeta'$ and vanishes in the 
de Sitter limit $\epsilon \to 0$.
One can verify that at linear order the combination
\ba
  \int\!\!\sqrt{-g} \left( R^{a}_{\,\,\, b}R^{b}_{\,\,\, a} - \frac{1}{3} R^2 \right) &=& 
  4 \int\!\!d^4x \left\{\partial_\tau( {\rho'}^{\, 3})  -  {\rho'}^{\, 3} \epsilon \zeta' + O(\zeta^2) \right\}
\ea
is in agreement with the identity 
\ba
  {\rm E}_4 =  C^2 - 2\left( R^{a}_{\,\,\, b}R^{b}_{\,\,\, a} - \frac{1}{3} R^2 \right)
\ea
which follows from (\ref{Riem squared}) below.
The last geometrical counterterm which is fourth order in the derivatives 
is $\int\!\!\sqrt{-g} \, \Box R$.
It is a bit ambiguous since it is a total derivative. 
If we drop the spatial divergences alltogether and expand the remainder at linear order,
we obtain
\ba
  \int\!\!\sqrt{-g} \, \Box R &=& \int\!\!d^4x\, \partial_a\left(  \sqrt{-g} g^{ab} \partial_b R \right)
\nonumber \\
  &=& - \int\!\!d^4x \, \partial_\tau\left( \frac{a^2 e^{3\zeta}}{N} R' \right)
\nonumber \\
  &=&  - 24  \int\!\!d^4x \, \epsilon^2 {\rho'}^3 \zeta'
\ea
It is subleading in the first slow roll parameter. 
If we do not drop the spatial derivatives, two additional linear terms
$\int\!\!d^4x \, e^\zeta a^2( \partial^2 R + R' \partial^2 \omega)$ are included.
We can integrate them by parts to produce a second order term 
$\propto \partial_i \zeta \partial_i \zeta'$.

Finally, it is easy to show that 
the counterterm \Eqref{cnt R^2} is comprised of 
all the terms at most quadratic in $\zeta$ with 
a total of four derivatives and a maximum number of two time derivatives of $\zeta$. 
Consider first 
\ba \label{Ricci squared}
  R^{a}_{\,\, b} R^{b}_{\,\, a} &=& (R^{0}_{\,\, 0})^2 + 2R^{i}_{\,\, 0} R^{0}_{\,\, i} 
  + R^{i}_{\,\, j} R^{j}_{\,\, i} 
\nonumber \\
  &=& (R^{0}_{\,\, 0})^2  + 3  \bar {\cal R}^2 + 2 \bar {\cal R} \delta^{ij} \delta_2 R_{ij} + O(\zeta^3)
\ea
The product of the $0$-$i$ terms is third order because 
$R^{i}_{\,\, 0} \propto \partial_i \omega$ and $R^{0}_{\,\, i} = 2\bar {\cal R}(\alpha + \zeta) \partial_i\omega$,
and the linear term of $R^{i}_{\,\, j} = \bar {\cal R} \delta_{ij} + \delta_2 R_{ij}$ vanishes.
We see that at second order \Eqref{Ricci squared} differs from 
$R^2 = \left(R^{0}_{\,\, 0} + 3  \bar {\cal R} + \delta^{ij} \delta_2 R_{ij}\right)^2$ 
only by a different combination of the same terms.
Similarly for 
\ba \label{Riem squared}
  R^{abcd}R_{abcd} = C^2 + 2 R^{ab}R_{ab} - \frac{1}{3} R^2
\ea
which follows from \Eqref{def Weyl}.
The last counterterm $\int \sqrt{-g} \, \Box R$ does not contain any new quadratic term either, 
since every time that a $\zeta''$ or 
$\sigma''$ appears we can substitute the linear field equations.

\section{Time integrals for a red spectrum}
\label{app:time int red spectrum}

Consider the model of power law inflation,
\ba \label{power law}
  a(\tau) &=& \left(\frac{\tau_0}{\tau} \right)^{1+\alpha}  
 \, , \qquad \partial_t^2 a > 0 \quad \Longleftrightarrow \quad \alpha > 0
\ea
The Hubble rate is $H(\tau) = H_0 \left( \frac{\tau}{\tau_0} \right)^\alpha$ 
with $H_0 = \frac{1+\alpha}{-\tau_0}$
and the only nonvanishing slow roll parameter is  
$\epsilon = \frac{\alpha}{1+ \alpha}$.
The solutions of the mode equations 
\ba 
  \zeta_{q}(\tau) &=& {\cal N}^{\zeta}_q \,
  \frac{\sqrt{-\tau}}{a(\tau)} {\cal H}_{\nu}^{(1)}(-q\tau)
\, , \qquad 
  \sigma_p(\tau) =  {\cal N}^{\sigma}_p \,
  \frac{\sqrt{-\tau}}{a(\tau)} {\cal H}_{\nu}^{(1)}(-p\tau)
\ea
have the same behaviour as (\ref{solzeta}-\ref{modemin}) for $q\tau \!\to -\!\infty$.
${\cal H}_{\nu}^{(1)}$ is the Hankel function of the first kind of index
\ba \label{scalefactor}
   \nu = \frac{3}{2} + \alpha >  \frac{3}{2}
\ea
I will assume that $\alpha$ does not take the values $\frac{1}{2}$ or 
$\frac{5}{2}$ for which the discussion
below must be amended. These special values correspond to 
$\epsilon = 1/3$ and $\epsilon = 5/7$ which are outside the slow-roll regime anyway.
To be precise, say $\alpha < \frac{1}{2}$ in the following.
Using the identity
$\frac{d{\cal H}_{\nu}}{dz} = {\cal H}_{\nu - 1}(z)  - \frac{\nu}{z}{\cal H}_{\nu}(z)$,
we have 
\ba
  \dot \zeta_q = -\frac{{\cal N}_q^{\zeta}}{a} \, \frac{q}{a} \, 
  \sqrt{-\tau}{\cal H}_{\nu-1}^{(1)}(-q\tau)
\ea
and a similar expression for $\dot \sigma_p$.
The normalization constants ${\cal N}^{\zeta}_q$ and ${\cal N}^{\sigma}_p$ are 
fixed by the canonical commutation
relations, e.g.
\ba
  \left[ \zeta_{\bf q}(t) ,\, \pi_{-{\bf q}}(t)  \right] = i &=& 
  \frac{a^3 \epsilon}{4\pi G} \left( \zeta_{q} \dot \zeta_{q}^* - c.c.  \right)
  \nonumber \\
  &=& \frac{a^3 \epsilon}{4\pi G} \frac{\abs{{\cal N}_q}^2}{a} 
 \left( z \frac{d{\cal H}_{\nu}^{(1)\, *}}{dz} \, {\cal H}_{\nu}^{(1)} - c.c. 
 \right)_{z=-q\tau}
\ea
The term in parenthesis is the Wronskian of the Hankel functions and is equal to
$-4i/(\pi z)$. We deduce
\ba
   \abs{{\cal N}^\zeta_q}^2 = \frac{\pi}{4} \frac{4\pi G}{\epsilon_q}
\, , \qquad 
   \abs{{\cal N}^\sigma_p}^2 = \frac{\pi}{4}  
\ea

To analyze the late time behaviour of the various integrals, 
I use the asymptotic expansion of the Hankels for small arguments
\ba \label{asympt}
  {\cal H}_{\nu}(z) \to \frac{-i}{\sin(\pi \nu)} \left[ 
  \frac{1}{\Gamma(1-\nu)} \left( \frac{2}{z} \right)^{\nu} - 
  e^{i\pi \nu} \frac{1}{\Gamma(1+\nu)} \left( \frac{z}{2} \right)^{\nu} \right]\left( 
  1+ O(z^2) \right)
\ea
which gives the following expansion of the mode
\ba \label{asympt mode}
  \zeta_q(\tau) = \zeta_q^0\left( 1 + i a_\nu (q\tau)^{3+2\alpha} + ... \right)
\ea
where $a_{\nu} = \vert a_\nu \vert \, \exp(i \pi \alpha)$.
The power spectrum is
\ba
  q^3 \vert \zeta_q(\tau) \vert^2 \to q^{-2\alpha} 
\ea
Given the condition \Eqref{power law}, this is a red spectrum
(more power in for large wavelengths).
One can explicitly check that this does not change the 
contribution of the vertex $\widetilde H_3$ of section \ref{sec:three vertex}:
\ba \label{compactpowerlaw}
  V_1 V_2 {\cal Z}_q^{\zeta \zeta}{\cal M}_{pp'}^{\sigma} &\sim&  
  \frac{\epsilon_q^2 H_0^2}{q^{6+4\alpha}} (-\tau_0)^{-2-4\alpha} \, 
  \frac{(pp')^{1-2\alpha}}{(\tau_1 \tau_2)^{2\alpha}}  
\left[ (-q\tau)^{3 + 2\alpha} + (-q\tau_2)^{1+2\alpha}  \right]
\ea
The limit $\alpha \to 0$ continuously converges to \Eqref{Sigma>} and \Eqref{Mred}.
The integrals over time of \Eqref{compactpowerlaw} converge for $\tau \to 0$.
By continuity, we can conclude that for small values of $\alpha$,
$\Delta {\cal P}_{\zeta}$ is given by its value \Eqref{univpower} plus a 
term $\propto \alpha \tau^{\alpha'}$ with $\alpha' > 0$ 
that goes to zero for $\tau \to 0$.

To evaluate the integrals (\ref{cnt zeta zeta'}), (\ref{cnt tadpole}), 
(\ref{formal tadpole}) and (\ref{formal delta_4 P}), we need too more ingredients.
One is the asymptotic expansion of $\zeta'/\rho'$
\ba
   \frac{\zeta_q'}{\rho'} = - (1+\alpha)q^2 \tau^2 \, \zeta_q^0\left( 1 + i b_\nu (q\tau)^{1+2\alpha} + ... \right) 
\ea
with the coefficient
\ba
  b_\nu = \vert b_\nu \vert e^{i\pi \alpha} \,, \qquad 
  \frac{\vert b_\nu \vert}{\vert a_\nu \vert } = (3+2\alpha)(1+2\alpha)
\ea
We also need the expression of the renormalized energy momentum tensor.
It can be found in \cite{BirrelDavis} 
as a polynome of $\rho'$ and its derivatives. I only quote the result
\ba \label{Tab power law}
  a^2\langle T_{00} \rangle &=& \frac{1}{960 \pi^2} \, \frac{1}{ \tau^4} \left( 119 + 482 \alpha + O(\alpha^2)   \right)
\nonumber \\
  a^2\langle T_{11} \rangle &=& - \frac{1}{960 \pi^2 } \, \frac{1}{\tau^4} \left( 119 + \frac{433}{3} \alpha + O(\alpha^2)   \right)
\ea
Because the scale factor is a monomial of $1/\tau$ in 
power law and de Sitter inflation, $\rho' \propto 1/\tau$ in both cases.
This means that the renormalized energy-momentum tensor contributes the same 
to the time integrals in both cases.

I now use these expressions to calculate the 
value of the tadpole (\ref{formal tadpole}). 
The terms $O(\alpha^0)$ of (\ref{Tab power law})
give back (\ref{zero tadpole}), and the terms  $O(\alpha)$ are multiplied by a factor
$\alpha^{-1}$ produced by the integration over time. The leading term of 
$\langle \zeta_{\bf q}(\tau) \rangle$ is therefore $O(\alpha^0)$.
The final result is
\ba
  \langle \zeta_{\bf q}(\tau) \rangle &=& \# \delta({\bf q}) \times q^3 (q\tau)^{\alpha} 
\ea
where $\#$ stands for a silly number. We have exactly the same result for (\ref{cnt tadpole}).

I then apply \Eqref{asympt mode} to the integral (\ref{cnt zeta zeta'}), 
and (\ref{delta_4 P}), which are also subject to identical changes.
For instance 
\ba \label{cnt zeta zeta' power law}
  Im \int^\tau\!\!d\tau_1\, {\rho'}^3  \langle  \zeta(\tau_1)\zeta'(\tau_1) \zeta^2(\tau) \rangle
  &\sim& q^{3}\vert \zeta_q^0 \vert^4 \,\times  \frac{1}{2\alpha} \left( \frac{q}{a\dot \rho} \right)^{2\alpha}
\ea
The r.h.s. is approximated by $\alpha^{-1} + \ln(-q\tau)$
for time intervals $q\tau \ll \alpha^{-1} \sim \epsilon^{-1}$
and asymptotes to zero. 
The large value for $q\tau \to 1$ is not significant since 
(\ref{cnt zeta zeta' power law}) is an asymptotic expansion.

\section{The CTP formalim}
\label{app:in-in}

For time dependent problems, where there is no stable ground state,
there exists a covariant formalism called the Closed Time Path formalism
adapted to calculate expectation values such as Green functions.
I briefly introduce this formalism for the readers convenience
and refer to \cite{CTP} for other short reviews and further references.
The important point for us is the structure of the one loop correction to
the two point function given in Sec. \ref{CTPstructure}.

\subsection{Motivation}
\label{CTPmotiv}

Let us consider the time-ordered two-point function, defined in the
Heisenberg picture by
\ba \label{exactGreen}
   G(x,y) = \bra{\Psi_0} {\cal T} \varphi_H(x)
   \varphi_H(y) \ket{\Psi_0} \, ,
\ea
where $\ket{\Psi_0}$ is the exact (non-degenerate) ground state 
at an 'initial' time $t_0$.
Inside the horizon, the evolution of the modes is adiabatic and 
a vacuum state is well defined.
The initial time can be taken $t_0 \to -\infty$ and the 
initial state in noted $\ket{\Psi_0} = \ket{0\, {\rm in}}$.

In the interaction picture, for the particular time ordering $x^0 > y^0$, 
Eq. \Eqref{exactGreen} becomes
\ba \label{identity1}
   G(x,y)  &=& \bra{0\, {\rm in}} 
    U_I(-\infty, x^0)\varphi_I(x) U_I(x^0,y^0)
   \varphi_I(y) U_I(y^0,-\infty) \ket{0\, {\rm in}}.
\ea
where $U_I$ is given by
\ba \label{U_I}
   U_I(t,t_0) = {\cal T} \exp\left( -i\int_{t_0}^{t}\!\!dt' \, H_I(t') \right)
\ea
and $H_I[\varphi_I,\pi_I;t]$ is the interaction Hamiltonian written in terms of the
field variables $\varphi_I,\pi_I$ in the interaction picture.
Let us insert a resolution of the identity at the time 
$t_{\rm out} = +\infty$ 
in the last line of \Eqref{identity1}
\ba \label{identity2}
  x^{0} > y^{0} \, , \quad 
  G(x,y)  &=& \sum_{n=0}^{\infty}
   \bra{0 \, {\rm in}} U_I(-\infty,+\infty) \ket{n\,{\rm out}}
\nonumber \\
   && \quad \times \, 
   \bra{n\,{\rm out}}U_I(+\infty,x^0)
   \varphi_I(x) U_I(x^0,y^0)
   \varphi_I(y) U_I(y^0,-\infty) \ket{0 \, {\rm in}}
\qquad
\ea
When the ground state is stable, the matrix element 
$\bra{0 \, {\rm in}} U_I(-\infty,+\infty) \ket{0\,{\rm out}}$ is a phase
and the matrix elements 
$\bra{0 \, {\rm in}} U_I(-\infty,+\infty) \ket{n\neq 0\,{\rm out}}$ vanish identically.
One thus recovers the familiar  
expression for the time ordered propagator in the interaction picture 
$G(x,y) = \bra{0 \, {\rm in}}{\cal T} U_I(\infty,+\infty)  \, 
 \varphi_I(x) \varphi_I(y)  \ket{0\,{\rm in}} / 
\bra{0 \, {\rm in}} U_I(+\infty,-\infty) \ket{0\,{\rm in}}$.
This simple expression no longer holds when the initial state is not stable
since then all the terms in the sum contribute.
In that case, the pertubative expansion of the Green function is obtained directly from 
\Eqref{identity1} after expanding each of 
the three evolution operators in \Eqref{U_I}. 
The perturbative theory can be described by a diagramatic
formalism akin to the Feynman rules for scattering matrix elements.

\subsection{The 'in-in' generating functional and Feynman rules}
\label{W}

The so-called 'in-in' generating functional of connected Green's functions 
is given by the transition amplitude of two 
'in'-vacua in the presence of external sources $J_+$ and $J_-$,
 \ba \label{Winin}
   e^{iW[J^+,\,J^-]} &\equiv&
   {}_{J_-} \langle 0 \, {\rm in} \vert
   0 \, {\rm in} \rangle_{J_+}
   \nonumber \\
   &=& \sum_{n=0}^{\infty}
   \bra{0 \, {\rm in}} \widetilde {\cal T} e^{-i\int_{-\infty}^{t_{out}}dt
   J_-(x) \varphi_H(x)}  \ket{n\,{\rm out}}
   \bra{n\,{\rm out}} {\cal T} e^{i\int_{-\infty}^{t_{out}}dt
   J_+(x) \varphi_H(x)}
   \ket{0 \, {\rm in}} 
\nonumber \\ 
   &=& {\rm Tr} \left\{ 
   {\cal T} e^{i\int_{-\infty}^{t_{out}}dt\, J_+(x) \varphi_H(x)} 
   \,  \rho_{\rm in} \, 
   \widetilde {\cal T} e^{-i\int_{-\infty}^{t_{out}}dt\, 
   J_-(x) \varphi_H(x)}  
   \right\}
\ea
where $\widetilde {\cal T}$ is the reversed-time ordered product and 
$J_+$ and $J_-$ are the two classical sources associated with the 
two branches of evolution, forward and backward in time respectively. 
Because of these two reverse branches, 
the formalism is called Close Time Path. 
In the second line the resolution of the indentity at an 
arbitrary time $t_{out}$ was inserted.
In the last line, the generating functional was written in 
a form that allows arbitrary initial states (not only the vacuum).  
Note that the operation of taking the trace
 couples the forward and backward time evolutions. 
As a result, $W[J_+,J_-]$ generates four types of connected 
two-point functions: the usual time-ordered propagator
\ba \label{G++}
   G_{++}(x,y) &\equiv& i\langle {\cal T} \varphi(x) \varphi(y)  \rangle = 
  \frac{\delta W}{\delta J_+(x) \delta J_+(y)}\vert_{J_+ = J_- =0}    
\ea
the reverse times ordered propagator
\ba \label{G--}
  G_{--}(x,y) &\equiv& i\langle \widetilde{\cal T} \varphi(x) \varphi(y)  \rangle =
  \frac{\delta W}{\delta J_-(x) \delta J_-(y)}\vert_{J_+ = J_- =0}  
\ea 
and the two on-shell two-point functions (the Wightman functions)
\ba \label{G-+}
   G_{-+}(x,y) &\equiv& i\langle \varphi(x) \varphi(y)  \rangle = 
  \frac{\delta W}{\delta J_-(x) \delta J_+(y)}\vert_{J_+ = J_- =0}  
\\ 
    \label{G+-}
   G_{+-}(x,y) &\equiv& i\langle \varphi(y) \varphi(x)  \rangle = 
  \frac{\delta W}{\delta J_+(x) \delta J_-(y)}\vert_{J_+ = J_- =0}    
\ea
The latters are not time ordered because
they are build with operators coming from different branches.

The path integral representation of the 'in-in' generating function 
is obtained from (\ref{Winin}) by doubling the fields, 
one $\varphi_+$ for the forward branch
and one $\varphi_-$ for the backward branch. The obtention of Feynman rules was described in 
details in \cite{CTP} and \cite{Weinberg}.
They are similar to the 'in-out' Feynamn rules, but for a few differences: diagrams are 
build from the four propragators \Eqref{G++}-\Eqref{G+-} and two types of vertices 
of polarity $+$ (the same as the 'in-out' vertices) or $-$ 
(with a relative minus sign). At each vertex, 
in addition to the conservation of momentum, 
one can attach only the propagators with end lines of the same polarity. 
For instance, we can have the sequence
$... \, G_{-+} ({\rm vertex} + )G_{+-} ({\rm vertex}  -) G_{--} \, ...$

All the algebraic theorems of the in-out formalism are naturally extended to 
the in-in formalism. 
In particular, the double Legendre transform of 
$W[J_+,J_-]$ gives the in-in generating functional
$\Gamma[\bar \varphi_+,\, \bar \varphi_-]$ of $1$-particle irreducible vertex functions. 
The second variational derivative of $\Gamma[\bar \varphi_+,\, \bar \varphi_-]$ 
therefore gives the matrix of self-energies:
\ba
  \Gamma[\bar \varphi_+,\, \bar \varphi_-] = S_2[\bar \varphi_+] - S_2[\bar \varphi_-]
 - \frac{1}{2}\int\!\!d^4x\int\!\!d^4y \, 
\left(
   \begin{array}{cc}
      \bar \varphi_+  & \bar \varphi_-        
   \end{array} 
   \right)
  \left(
   \begin{array}{cccc}
           \Sigma_{++} &  \Sigma_{+-}  \\
           \Sigma_{-+}  & \Sigma_{--}   
   \end{array} 
   \right)    
\left(
   \begin{array}{c}
          \bar \varphi_+  \\
          \bar \varphi_-       
   \end{array} 
   \right)
\qquad
\ea
where $S_2$ is the quadratic part of the classical action, e.g. \Eqref{S_2}.
The self-energy matrix elements are
\ba 
  \Sigma_{-+}(x,y)    &=& \Sigma_>(x,y)
\nonumber \\  
 \Sigma_{+-}(x,y)  &=& \Sigma_<(x,y)
\nonumber \\ 
  \Sigma_{++}(x,y) &=&  \theta(x^0 - y^0) \Sigma_>(x,y) + \theta(y^0 - x^0) \Sigma_>(x,y)
\nonumber \\ 
  \Sigma_{--}(x,y) &=&  \theta(x^0 - y^0) \Sigma_<(x,y) + \theta(y^0 - x^0) \Sigma_>(x,y)
\ea
The in-in effective action is renormalized by the same counterterms 
as the in-out effective action, because the on-shell self-energy 
(as well as higher order vertices) is finite and $\Sigma_{--}$
is the complex conjugate of time order self-energy $\Sigma_{++}$.

\subsection{Generic structure of the one-loop correction to the propagator from $H_3$}
\label{CTPstructure}

Consider for simplicity a theory with trilinear couplings $\varphi \psi^2$ between two
scalar fields.
Using the in-in Feynman rules, the one-loop correction to the Feynman propagator is
\ba \label{DeltaG_F}
   \Delta G_F(t, t',{\bf x}-{\bf y}) &=& 
  \int_{-\infty}^{t}\!\!d^4z_1 \int_{-\infty}^{t}\!\!d^4z_2 \,  
  G_{+a}^{\varphi}(x-z_1)\, \left[-i\Sigma_{ab}^{\psi}(z_1 - z_2)\right] \,  
  G_{b+}^{\varphi}(z_2 - y)
\qquad 
\ea
where the indices are summed with the metric $c_{++} = + 1$, $c_{--} = -1$, and
$c_{+-} = c_{-+} = 0$, 
and the matrix elements $G_{ab}$ are given at 
\Eqref{G++}-\Eqref{G+-}.
The self-energy matrix is 
\ba \label{Sigmamatrix}
  \Sigma_{ab}^{\psi}(z) = G_{ac}^{\psi}(z)   G_{cb}^{\psi}(z) 
\ea
The matrix proguct gives the following integrand 
\ba \label{intermed1}
  &&G^{++}(t,t_1) \left\{ \Sigma_{++} G^{11}(t_2,t) - \Sigma_{+-} G^{-+}(t_2,t) \right\} 
  - \left\{G^{+-}(t,t_1) \Sigma_{-+} - 
  G^{+-}(t,t_1) \Sigma_{--} \right\} G^{-+}(t_2,t) 
\nonumber \\
  &=& G_>(t,t_1)\left\{  \Sigma_{++} G_<(t_2,t) - G_>(t_2,t) \Sigma_< \right\} - 
 G_<(t,t_1)\left\{ \Sigma_{>}  G_<(t_2,t) - \Sigma_{--}  G_>(t_2,t)    \right\}
\nonumber 
\ea
To get the second line, I expanded the time ordered propagators, and anticipating on
the integration over time I dropped the
terms proportional to $\theta(t_1 - t)$ and $\theta(t_2 - t)$.
Subtituting these expressions into \Eqref{intermed1}, one finally obtains
\ba
    \Delta G_F(t,t,{\bf x}-{\bf y}) &=& 
 {\rm Re} \int_{-\infty}^{t}\!\!dz_2^0 \int_{-\infty}^{z_2^0}\!\!dz_1^0 \,  
  \left( G_>(t',z_2^0) -  G_<(t',z_2^0) \right) \,
\nonumber \\ 
  &&\qquad 
 \times \, \left[  G_>(t,z_1^0) \Sigma_>(z_0^1,z_0^2) - G_<(t,z_1^0) \Sigma_<(z_0^1,z_0^2)
\right]
\nonumber \\
\ea

\end{appendix}

\end{document}